\documentclass[reprint,superscriptaddress,amsmath,amssymb,aps,onecolumn]{revtex4-1}
\usepackage{graphicx,graphics}
\usepackage{dcolumn}
\usepackage{bm}
\usepackage{hyperref}
\usepackage{enumitem}
\usepackage{color}
\usepackage{setspace}

\usepackage{lineno}
\newcommand*\patchAmsMathEnvironmentForLineno[1]{%
  \expandafter\let\csname old#1\expandafter\endcsname\csname #1\endcsname
  \expandafter\let\csname oldend#1\expandafter\endcsname\csname end#1\endcsname
  \renewenvironment{#1}%
     {\linenomath\csname old#1\endcsname}%
     {\csname oldend#1\endcsname\endlinenomath}}% 
\newcommand*\patchBothAmsMathEnvironmentsForLineno[1]{%
  \patchAmsMathEnvironmentForLineno{#1}%
  \patchAmsMathEnvironmentForLineno{#1*}}%
\AtBeginDocument{%
\patchBothAmsMathEnvironmentsForLineno{equation}%
\patchBothAmsMathEnvironmentsForLineno{align}%
\patchBothAmsMathEnvironmentsForLineno{flalign}%
\patchBothAmsMathEnvironmentsForLineno{alignat}%
\patchBothAmsMathEnvironmentsForLineno{gather}%
\patchBothAmsMathEnvironmentsForLineno{multline}%
}

\begin{document}

\title{Effects of Isothermal Stratification Strength on Vorticity Dynamics for Single-Mode Compressible Rayleigh-Taylor Instability}

\author{Scott A. Wieland}
\email{scott.wieland@colorado.edu}
\affiliation{Department of Mechanical Engineering, University of Colorado, Boulder}

\author{Peter E. Hamlington}
\email{peh@colorado.edu}
\affiliation{Department of Mechanical Engineering, University of Colorado, Boulder}

\author{Scott J. Reckinger}
\email{scotreck@uic.edu}
\affiliation{Department of Mechanical \& Industrial Engineering, University of Illinois Chicago}

\author{Daniel Livescu}
\email{livescu@lanl.gov}
\affiliation{Los Alamos National Laboratory, Los Alamos, New Mexico}

\date{\today}

%%%%%%%%%%%%%%%%%%%%%%%%%
%% Abstract
%%%%%%%%%%%%%%%%%%%%%%%%%
\begin{abstract} 
The effects of isothermal stratification strength on vorticity dynamics for single-mode Rayleigh-Taylor instability (RTI) are examined using two dimensional fully compressible wavelet-based direct numerical simulations. The simulations model low Atwood number ($A=0.04$) RTI development for four different stratification strengths, corresponding to Mach numbers from 0.3 (weakly stratified) to 1.2 (strongly stratified), and for three different perturbation Reynolds numbers, from 5,000 to 20,000. All simulations use adaptive wavelet-based mesh refinement to achieve very fine spatial resolutions at relatively low computational cost. For all stratifications, the bubble and spike go through the exponential growth regime, followed by a slowing of the RTI evolution. For the weakest stratification, this slow-down is then followed by a re-acceleration, while for stronger stratifications the suppression of RTI growth continues. Bubble and spike asymmetries are observed for weak stratifications, with bubble and spike growth rates becoming increasingly similar as the stratification strength increases. For the range of cases studied, there is relatively little effect of Reynolds number on bubble and spike heights, although the formation of secondary vortices becomes more pronounced as Reynolds number increases. The underlying dynamics are analyzed in detail through an examination of the vorticity transport equation, revealing that incompressible baroclinicity drives RTI growth for small and moderate stratifications, but increasingly leads to the suppression of vorticity production and RTI growth for stronger stratifications. These variations in baroclinicity are used to explain the suppression of RTI growth for strong stratifications, as well as the anomalous asymmetry in bubble and spike growth rates for weak stratifications.
\end{abstract}


\maketitle

%%%%%%%%%%%%%%%%%%%%%%%%%
%% Introduction
%%%%%%%%%%%%%%%%%%%%%%%%%
\section{Introduction\label{sec:intro}}

%Setup and motivate the physical goals of this paper
Rayleigh-Taylor instability (RTI) is formed at the interface of two fluids with different densities when an accelerative force is applied across the interface in the direction of the less dense fluid \cite{Rayleigh1884,Taylor1950}. Such a scenario arises in a number of engineering and physics problems, including inertial confinement fusion (ICF) \cite{Betti1998,Craxton2015}, supernova ignition fronts \cite{Hachisu1991,Zingale2005,Cabot2006,JordanIV2008,Swisher2015}, X-ray bursts \cite{Lewin1997}, and various topics in geophysics \cite{Sharp1984,Houseman1997,Lawrie2011}, to name just a few examples. In each of these cases, both flow and fluid compressibility may affect RTI growth; the former is related to the thermodynamic state and the stratification of background density and pressure fields, while the latter is related to the equation of state and differences in the specific heat ratio between the two fluids \cite{Livescu2004} (Ref.\ \cite{Gauthier2017} refers to these two effects as ``static'' and ``dynamic'' compressiblity, respectively). Moreover, in many of these cases, as well as in many experiments \cite{Dalziel1999,Ramaprabhu2004,Andrews2010,Lawrie2011,Zhou2017}, the Atwood number (namely, the density ratio of the two fluids normalized to take values between 0 and 1) is small and the Reynolds number is large, resulting in compressible dynamics driven by relatively small density differences over a wide range of length and time scales. In the present study, two dimensional (2D) fully compressible wavelet-based direct numerical simulations (DNS) are used to examine, from a dynamical standpoint, the evolution of low Atwood number RTI for different isothermal stratification strengths (i.e., flow compressibility) and Reynolds numbers. 

%Make the caveat that ours is not the first to do fully compressible DNS, but we are unique
Until relatively recently, many DNS studies of RTI used incompressible, low-Mach number, Boussinesq, or anelastic approximations to reduce the computational cost, often yielding valuable physical insights (see, e.g., \cite{Cook2001,Livescu2004,Livescu2008,Livescu2009,Schneider2016,Schneider2016b}). However, the present study is one of a growing number of fully compressible DNS analyses of RTI growth and characteristics. Lafay \emph{et al.} \cite{Lafay2007} examined RTI growth in the linear regime for different compressibility strengths (addressing both flow and fluid compressibility), and Gauthier \cite{Gauthier2013} examined RTI growth into the nonlinear regime for two different stratification strengths. More recently, Reckinger \emph{et al.} \cite{Reckinger2016} examined single-mode, 2D RTI growth rates for a range of stratification strengths, and Gauthier \cite{Gauthier2017} performed a comprehensive study of the dynamics of multi-mode, three-dimensional (3D) RTI for a relatively strongly stratified case. Both of these more recent studies employed variable-resolution numerical methods to achieve high Reynolds numbers within the context of fully compressible DNS; Reckinger \emph{et al.} \cite{Reckinger2016} used the parallel adaptive wavelet collocation method (PAWCM) \cite{Reckinger2010} and Gauthier \cite{Gauthier2017} used an auto-adaptive multi-domain Chebyshev-Fourier method \cite{LeCreurer2008}. Using currently available computational resources, these and other adaptive techniques are unavoidable when performing fully compressible DNS at high Reynolds numbers. The present study correspondingly employs PAWCM to study the effects of flow compressibility and Reynolds number, including high Reynolds numbers, on RTI growth and dynamics. 

%Outline prior stratification strength results
Based in large part on observations from these prior fully compressible DNS studies, the general effects of compressibility on RTI growth are now relatively well understood. Despite some initial ambiguity regarding the specific impacts of compressibility (dating back, at least, to the studies by Bernstein \& Book \cite{Bernstein1983} and Baker \cite{Baker1983}), Livescu \cite{Livescu2004} used a linear analysis of the Navier-Stokes equations to show that, for isothermal background stratification, flow compressibility is associated with a reduction in the rate of RTI growth, while fluid compressibility is associated with an increase in the growth rate, compared to the corresponding incompressible case. A number of studies have confirmed these results, particularly with respect to the suppression of RTI growth by flow compressibility \cite{Lafay2007,Yu2008,Gauthier2010,Wei2012,Reckinger2016,Gauthier2017}. In particular, as the stratification strength of the background density field increases, an increasing suppression of RTI growth has been observed. Reckinger \emph{et al.} \cite{Reckinger2016} further found that there are asymmetries in the locations and speeds of upward propagating low density fluid (i.e., ``bubbles'') and downward propagating high density fluid (i.e., ``spikes''), even at relatively small Atwood number, that may be different than in the incompressible limit. It was also shown by Reckinger \emph{et al.} \cite{Reckinger2016} that drag and potential flow models are unable to predict the suppression of RTI growth for strong stratifications.

%Outline prior Reynolds number results
Compared to the effects of compressibility, Reynolds number effects on RTI growth have received somewhat less attention. Wei \& Livescu \cite{Wei2012} used the incompressible variable-density form of the Navier-Stokes equations to show that, at early times, RTI growth rates are larger for smaller Reynolds numbers, due to diffusive effects, but that at long times, RTI growth rates are greater for larger Reynolds numbers. For sufficiently high Reynolds numbers, Wei \& Livescu \cite{Wei2012} found that, at low Atwood numbers, bubble and spike growth rates become quadratic, similar to the multi-mode case and contrary to the ubiquitous ``terminal  velocity'' assumption in previous studies. The single-mode growth rate ceases to depend on the Reynolds number at sufficiently large Reynolds numbers and may constitute an upper limit for the multi-mode growth rate. Using fully compressible DNS, Gauthier \cite{Gauthier2017} similarly found that smaller Reynolds numbers are associated with faster early growth rates of the turbulent mixing layer produced by the RTI. At later times, growth rates for higher Reynolds numbers are similar to, or exceed, those of lower Reynolds numbers. These results were, however, obtained for a single stratification strength, and it remains to be seen how these Reynolds number effects depend on stratification strength, if at all. 

%Outline prior vorticity results
In order to understand compressibility and Reynolds number effects in more detail, several authors have examined the dynamics of the vorticity during RTI evolution, generally finding that changes in the baroclinic torque are responsible for changes in RTI growth rates. Lafay \emph{et al.} \cite{Lafay2007} examined the linear regime and found that vorticity production decreases as the stratification strength increases. More recently, Schneider and Gauthier \cite{Schneider2016a} performed a systematic study of vorticity during RTI growth using 3D multimode simulations that employ the Boussinesq approximation. This study showed that there is an increase in the strength of baroclinic torque production with time, although the contribution to the overall dynamics is dwarfed by the effects of nonlinear vortex stretching. Gauthier \cite{Gauthier2013,Gauthier2017} was the first to examine vorticity using fully compressible DNS, and showed the importance of baroclinic torque in producing vorticity during RTI growth for a single strongly stratified case. However, changes to the relative magnitudes of the various terms in the vorticity transport equation for different stratification strengths are still not completely understood in the fully compressible case.  
       
%Questions to be addressed in this study and novelty
Despite the improved understanding of compressibility and Reynolds number effects provided by the recent, primarily computational, studies noted above, a number of outstanding questions remain, and the present study is specifically focused on addressing the following: (\emph{i}) How does the behavior of low Atwood number RTI depend on both stratification strength and Reynolds number?; (\emph{ii}) What are the dynamical causes of the observed RTI phenomena?; and (\emph{iii}) How do the dynamics (specifically, the vorticity dynamics) depend on stratification strength? The first question is motivated by the studies of Lafay \emph{et al.} \cite{Lafay2007} and Wei \& Livescu \cite{Wei2012}; the former studied the effects of compressibility, but within the linear regime and for only one Reynolds number, while the latter studied a range of Reynolds numbers, but using an incompressible variable-density formulation of the Navier-Stokes equations that precluded the study of compressibility effects. The second and third questions are motivated primarily by the studies of Reckinger \emph{et al.} \cite{Reckinger2016}, Schneider \& Gauthier \cite{Schneider2016a}, and Gauthier \cite{Gauthier2017}. The first of these studies observed bubble-spike asymmetries but not their dynamical causes, while the second and third studies both performed extensive analyses of the vorticity dynamics, but using the Boussinesq approximation (i.e., not a fully compressible study) and for only one stratification strength, respectively.

%Provide an overview of our approach
In the present paper, DNS are performed at low Atwood numbers (0.04 here, as compared to 0.1-0.7 in \cite{Reckinger2016}) for different Reynolds numbers and different strengths of initial hydrostatic stratification, corresponding to Mach numbers between 0.3 (weak stratification) and 1.2 (strong stratification) \cite{Livescu2013,Livescu2004,Gerashchenko2016}. The DNS are performed using adaptive mesh refinement based on PAWCM, as described, validated, and implemented for RTI by Reckinger \emph{et al.} \cite{Reckinger2010,Reckinger2016}. This method allows high spatial resolution to be used where it is needed (e.g., where density and velocity gradients are large), while reducing the total number of computational collocation points. The present focus on low Atwood numbers is motivated primarily by the observations of quadratic high Reynolds number single-mode RTI growth in regimes with similarly low Atwood number in the study by Wei \& Livescu \cite{Wei2012}. In order to understand the dynamical causes of the observed results, the various terms in the vorticity transport equation are examined as functions of time and stratification strength.

%Simplifications made to this problem
It should be noted that several simplifications are made here to allow the underlying physics to be more easily understood. In particular, complex interactions of multiple wavelengths are eliminated by applying only single-mode initial perturbations to the unstable interface between the two fluids with differing densities. Moreover, in the classical incompressible case, where the density of both fluids is constant, RTI growth eventually leads to a re-acceleration of the bubble and spike tips, finally resulting in chaotic dynamics and development. In the compressible case, however, matters can become more complicated due to spatial and temporal variations in the background density, pressure, and temperature fields. The effects of changing any of these fields are largely unknown, and thus only isothermal initial stratifications are studied here to eliminate thermal effects, since the initial state is already in thermal equilibrium. Future work will explore the effects of multi-modal perturbations and different stratification types.  Finally, the present simulations and analysis are performed in 2D in order to enable the examination of several different stratification strengths and Reynolds numbers. Each such simulation is computationally expensive and performing a similarly expansive study in 3D remains the focus of future research, due primarily to the need for substantially more extensive computational resources. The primary disadvantage of the present 2D approach is the resulting lack of nonlinear vortex stretching in the vorticity dynamics, although the absence of this effect does have the benefit of more clearly revealing the effects of baroclinicity on the dynamics.   

%Outline the paper
The rest of this paper is organized as follows. The next section discusses the problem setup, including the governing equations and initialization of the RTI. Section \ref{sec:method} provides a brief discussion of how the wavelet-based adaptive method (i.e., PAWCM) was used to complete the simulations. In Section \ref{sec:results}, the paper goes in depth into the results of this study, looking at the effects of stratification strength and Reynolds number on RTI growth. In Section \ref{sec:vorticity}, the dynamics of the vorticity for fully compressible RTI are outlined and examined. Finally, a summary and conclusions are presented in Section \ref{sec:conclusions}.

%%%%%%%%%%%%%%%%%%%%%%%%%
%% Problem description
%%%%%%%%%%%%%%%%%%%%%%%%%
\section{Description of the Physical Problem\label{sec:problem}}
In the present study, RTI occurs through the initial placement of a heavier fluid, denoted by index `2' with molar mass $W_2$, above a lighter fluid, denoted by index `1' with molar mass $W_1$, in the presence of a gravitational accelerative force. The addition of a perturbation causes the pseudo-stable initial condition to be lost, and the heavier fluid begins to fall into the lighter fluid in a spike-like formation, while the lighter fluid rises into the heavier fluid in a bubble-like formation. In the following, the fully compressible fluid flow equations solved by the DNS are outlined, followed by a description of the initial isothermal hydrostatic stratifications of different strengths (as characterized by a static Mach number).

%=====================================
% Governing equations
%=====================================
\subsection{Governing Equations\label{subsec:goveqs}}
The numerical simulations solve the fully compressible Navier-Stokes equations for two miscible fluids given by \cite{Livescu2013}
\begin{align}
  \label{e:continuity}
  \frac{\partial\rho}{\partial t} + \frac{\partial(\rho u_j)}{\partial x_j} &= 0\,,\\
  \label{e:navier}
  \frac{\partial(\rho u_i)}{\partial t} + \frac{\partial(\rho u_i u_j)}{\partial x_j} &= -\frac{\partial p}{\partial x_i} + \rho g_i + \frac{\partial \tau_{ij}}{\partial x_j}\,, \\
  \label{e:energy}
  \frac{\partial(\rho e)}{\partial t} + \frac{\partial(\rho e u_j)}{\partial x_j} &= -\frac{\partial (p u_i)}{\partial x_i} + \rho u_i g_i + \frac{\partial (\tau_{ij} u_i)}{\partial x_j} - \frac{\partial q_j}{\partial x_j} + \frac{\partial [T(c_{\mathrm{p}})_l s_{jl}]}{\partial x_j}\,, \\
  \label{e:species}
  \frac{\partial(\rho Y_i)}{\partial t} + \frac{\partial(\rho Y_i u_j)}{\partial x_j}& = \frac{\partial s_{ji}}{\partial x_j}\,,
\end{align}
where $\rho$ is the density, $u_i$ is the velocity in the $x_i$ direction, $p$ is the pressure, $g_i$ is the gravitational acceleration, $\tau_{ij}$ is the viscous stress tensor, $e$ is the specific total energy, $q_i$ is the heat flux, $T$ is the temperature, $(c_\mathrm{p})_l$ is the specific heat capacity at constant pressure for fluid $l$, $s_{ji}$ is the mass flux for fluid $i$ in the $x_j$ direction, and $Y_i$ is the mass fraction for the $i^\textrm{th}$ fluid. Note that, for a two-fluid system, $Y_2=1-Y_1$, and so Eq.\ (\ref{e:species}) is only solved in the present simulations for $i=2$ (i.e., the heavier fluid). The pressure and caloric ideal gas laws are assumed to hold, so that the pressure and specific total energy can be expressed as
\begin{eqnarray}
p&=&\rho R T\,, \\
  e &=& \frac{1}{2} u_i u_i + c_\mathrm{v} T\,,   \label{e:e}\end{eqnarray}% 
where $R$ is the mixture gas constant defined in terms of the universal gas constant $\mathcal{R}$ and the molar mass of each fluid, $W_i$, as
\begin{equation}
\label{e:R}
R = Y_i R_i =\mathcal{R}\frac{Y_i}{W_i}\,.
\end{equation}
In the above expression, the species gas constant is defined as $R_i\equiv \mathcal{R}/W_i$. The mixture specific heat at constant volume, $c_\mathrm{v}$, appearing in Eq.\ (\ref{e:e}) is similarly defined as
\begin{equation}
\label{e:cp}
c_\mathrm{v} = (c_\mathrm{v})_i Y_i\,,
\end{equation}
where the specific heats at constant pressure and volume are related by $(c_\mathrm{p})_i =(c_\mathrm{v})_i +R_i$ and their mixture values by $c_\mathrm{p}=c_\mathrm{v}+R$. The specific heats at constant volume are assumed constant and the same for the two fluids, so that the mixture specific heat at constant pressure varies with the flow due to the different molar masses of the two fluids. 

The viscous stress $\tau_{ij}$ in Eqs.\ (\ref{e:navier}) and (\ref{e:energy}) is assumed to be Newtonian and is given by
\begin{equation}
  \label{e:tau}
  \tau_{ij} = \mu\left(\frac{\partial u_i}{\partial x_j} + \frac{\partial u_j}{\partial x_i} - \frac{2}{3}\frac{\partial u_k}{\partial x_k}\delta_{ij}\right)=2\mu S'_{ij}\,,
\end{equation}% 
where $S'_{ij} = S_{ij} - S_{kk}\delta_{ij}/3$ is the deviatoric strain rate and the dynamic viscosity is given by $\mu = \rho \nu$, with the kinematic viscosity $\nu$ assumed to be constant (i.e., temperature independent and the same for both fluids) such that spatial and temporal variations in $\mu$ are due entirely to variations in $\rho$. The strain rate tensor $S_{ij}$ is given by
\begin{equation}
S_{ij} = \frac{1}{2}\left( \frac{\partial u_i}{\partial x_j} + \frac{\partial u_j}{\partial x_i}\right)\,.
\end{equation}
The heat flux in Eq.\ (\ref{e:energy}) is written as
\begin{equation}
  \label{e:q}
  q_j = -k\frac{\partial T}{\partial x_j}\,,
\end{equation}%
where $k$ is the thermal conductivity, and the species mass flux in Eqs.\ (\ref{e:energy}) and (\ref{e:species}) is defined as
\begin{equation}
  \label{e:s}
  s_{ji} = \rho D \frac{\partial Y_i}{\partial x_j}\,,
\end{equation}%
where $D$ is the mass diffusivity. For the range of parameters considered here, the baro-diffusion term is small in the mass flux, and Soret and Dufour effects are neglected in the mass and heat fluxes, respectively.  Both $k$ and $D$, like the kinematic viscosity $\nu$, are assumed to be constant and temperature-independent, and both Prandtl and Schmidt numbers are unity. 

The majority of fluid properties are taken to be the same between the two fluids for simplicity. This includes the kinematic viscosity, $\nu$, the heat conduction coefficient, $k$, and the mass diffusion coefficient, $D$. It should be noted that effects due to bulk viscosity and non-equilibrium thermodynamics are neglected in the simulations. Investigating these effects is beyond the scope of the present study, although Sagert \emph{et al.} \cite{Sagert2015} and Lai \emph{et al.} \cite{Lai2016}  have recently made progress in this direction.

The system of equations given by Eqs.\ (\ref{e:continuity})-(\ref{e:s}) is solved using the PAWCM numerical approach, which is described in Section \ref{sec:method}, for an RTI with a physical setup as outlined in the following section. 

%=====================================
% Initialization
%=====================================
\subsection{Initialization of Rayleigh Taylor Instability\label{subsec:rti}}
The RTI problem is initialized in the DNS by imposing a perturbation on a stratified isothermal background state that is in hydrostatic equilibrium. The gravitational acceleration is assumed to be in the negative $x_1$ direction, such that $g_i = -g \delta_{i1}$, where $g$ is the magnitude of the gravitational acceleration. The resulting density, $\rho(x_1,x_2,t)$, and pressure, $p(x_1,x_2,t)$, fields at $t=0$ can be expressed as
\begin{align}
\rho(x_1,x_2,0) & = \rho_0(x_1)+\rho'(x_1,x_2,0)\,, \\
p(x_1,x_2,0) & = p_0(x_1)+p'(x_1,x_2,0)\,,
\end{align}
where $\rho_0$ and $p_0$ are hydrostatic initial background states and $\rho'(x_1,x_2,0)$ and $p'(x_1,x_2,0)$ represent the initial perturbations to the background states.

Assuming an isothermal background state at temperature $T_0$, the background density and pressure fields for fluid $\alpha$ (where $\alpha=[1,2]$ and summation over Greek indices is not implied) are given by
\begin{align}
 \label{e:rhootherm}
  \rho_{0\alpha}(x_1) &= \frac{p_{I}}{R_\alpha T_0} \exp\left(-\frac{g x_1}{R_\alpha T_0}\right)\,,\\
  \label{e:pisotherm}
  p_{0\alpha}(x_1) &= p_{I} \exp\left(-\frac{g x_1}{R_\alpha T_0}\right)\,,
\end{align}%
where the initial interface between the two fluids lies at $x_1=0$, $p_{I}$ is the interfacial pressure, and $R_\alpha=\mathcal{R}/W_\alpha$ is the gas constant based on the molar mass of fluid $\alpha$. The heavier fluid ($\alpha=2$) is initially located above the interface for $x_1>0$ and the lighter fluid ($\alpha=1$) is initially located below the interface for $x_1<0$. A corresponding interfacial density is given using the ideal gas law as $\rho_{I} = p_{I}/(R_I T_0)$ where $R_I = \mathcal{R}[(W_1+W_2)/2]^{-1}$.  

In each of the cases examined here, the kinematic viscosity $\nu=\mu/\rho$, which is constant and the same in both fluids, is set using the perturbation Reynolds number, $Re_p$, as
\begin{equation}
  \label{e:rey}
  Re_p \equiv \sqrt{\frac{A g \lambda^3}{(1+A)\nu^2}}\quad \Rightarrow \quad \nu = \sqrt{\frac{A g \lambda^3}{(1+A)Re_p^2}}\,,
\end{equation}%
where $\lambda$ is the wavelength of the applied perturbation and $A$ is the non-dimensional Atwood number defined as
\begin{equation}
  \label{e:A}
  A\equiv \frac{W_2 - W_1}{W_2 + W_1}\,.
\end{equation}%
Note that in the present study, $W_2 > W_1$ in order to generate RTI. 

The degree of flow compressibility defined by the thermodynamic conditions enters the RTI problem by affecting both the background stratification and the further development of dilatational (non-zero velocity divergence) effects \cite{Livescu2004,Livescu2013}. While dilatational effects and their acoustic manifestations are usually characterized by the Mach number denoting the ratio between velocity and sound speed, together with dilatational to solenoidal kinetic energy ratios, stratification strength can also be recast as a Mach number. This can be done by re-expressing $g x_1/(R_\alpha T_0)$ in Eqs.\ (\ref{e:rhootherm}) and (\ref{e:pisotherm}), as described below. 

In the present study, the relevant incompressible limit is found by simultaneously increasing the background pressure and temperature to cause an increase in the speed of sound such that the density remains unaffected. This incompressible limit is also easily obtained in practice by uniformly heating a fixed volume of fluid. This results in the definition of an isothermal Mach number based on the ratio of the gravity wave speed,  $\sqrt{g\lambda}$, and the isothermal speed of sound, $a_0 = \sqrt{{p_I}/{\rho_I}}$ \cite{Livescu2004,Yu2008}. The resulting Mach number, $M$, is then given by 
\begin{equation}
  \label{e:M}
  M = \sqrt{\frac{\rho_I g \lambda}{p_I}}\quad \Rightarrow \quad M^2=\frac{g\lambda}{R_I T_0}\,.
\end{equation}%
It should be noted that $M$ is equivalent to stratification strength parameters used in prior studies of flow (or ``static'') compressibility \cite{Schneider2016,Gauthier2017,Gauthier2017a}, and that larger values of $M$ indicate stronger stratification. 

\begin{figure}[t]
\centering
 \includegraphics[scale=0.9]{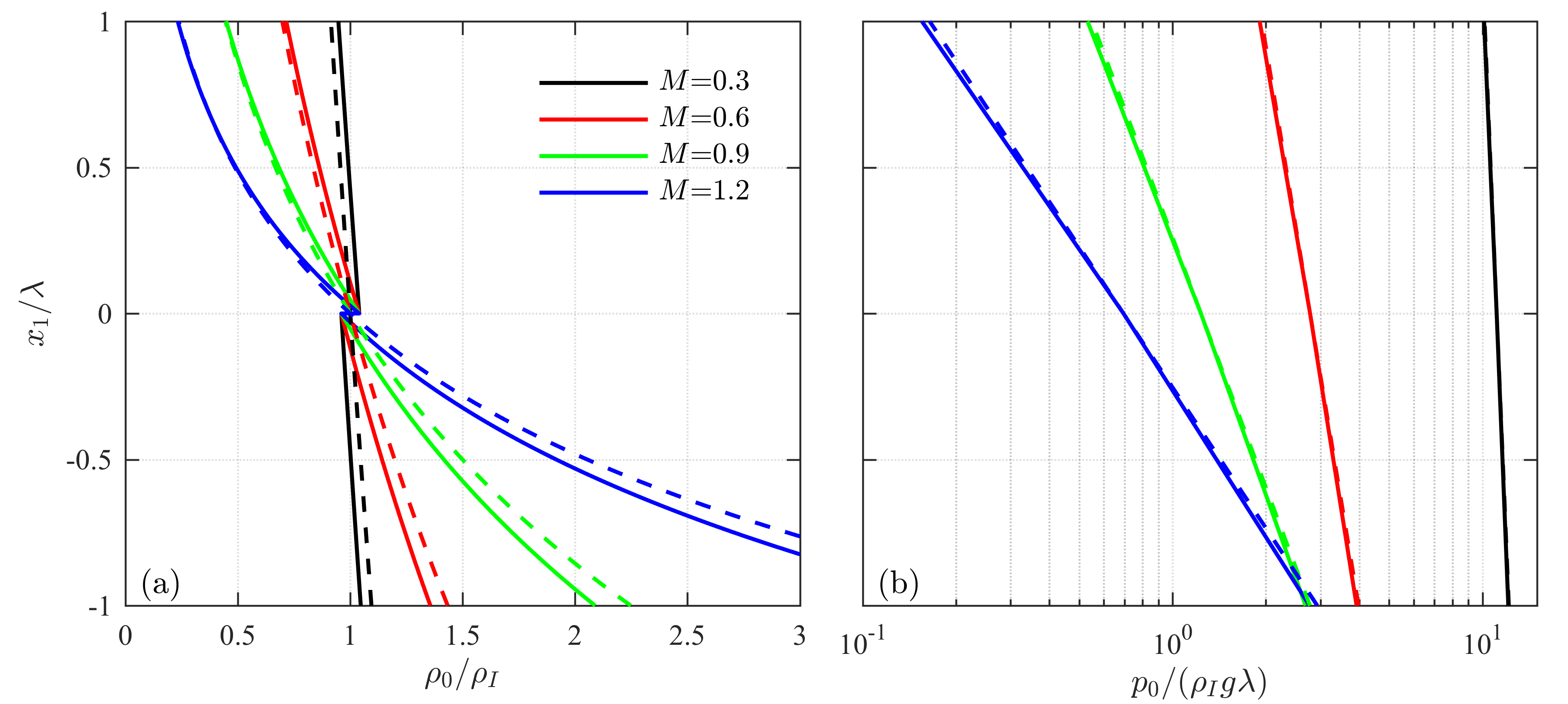}
\singlespacing{ \caption{[Color online] Background density (a) and pressure (b) profiles for $A=0.04$ and stratification strengths from $M=0.3$ to $1.2$. The background states, indicated by solid lines, are hydrostatic and are given by Eqs.\ (\ref{e:rhoothermfinal}) and (\ref{e:pisothermfinal}). The density difference at $x_1=0$ is determined by $A$. The dashed lines show the $A=0$ background profiles used in Section \ref{sec:vorticity} for the analysis of baroclinic torque in the vorticity equation. \label{f:init}}}
\end{figure}

Normalizing  $\rho_{0\alpha}$ in Eq.\ (\ref{e:rhootherm}) and $p_{0\alpha}$ in Eq.\ (\ref{e:pisotherm}) by $\rho_I$, $g$, and $\lambda$, the non-dimensional background states can be rewritten as
\begin{align}
  \label{e:rhoothermnorm}
  \rho^*_{0\alpha}(x^*_1) &= \frac{R_I}{R_\alpha} \exp\left(-M^2 \frac{R_I}{R_\alpha}x^*_1 \right)\,,\\
  \label{e:pisothermnorm}
  p^*_{0\alpha}(x^*_1) &= \frac{1}{M^2} \exp\left(-M^2 \frac{R_I}{R_\alpha}x^*_1 \right)\,,
\end{align}%
where the characteristic pressure is given as $\rho_I g \lambda$ and $x^*_1\equiv x_1/\lambda$ is a normalized distance variable. It can be shown that the ratio $R_I/R_\alpha$ can be written in terms of the Atwood number $A$ as
\begin{equation}
\frac{R_I}{R_\alpha} = \frac{2W_\alpha}{W_1+W_2} = 1+ (-1)^\alpha A \textrm{ for } \alpha=1,2\,.
\end{equation}
Since $\alpha=1$ corresponds to the lighter fluid for which $x^*_1<0$ initially and $\alpha=2$ corresponds to the heavier fluid for which $x^*_1>0$, the non-dimensional background states $\rho^*_0$ and $p^*_0$ can be written in final form as
\begin{align}
  \label{e:rhoothermfinal}
  \rho^*_{0}(x^*_1) &= \left(1 \pm A\right) \exp\left[-M^2 (1 \pm A) x^*_1\right]\,,\\
  \label{e:pisothermfinal}
  p^*_{0}(x^*_1) &= \frac{1}{M^2}\exp\left[-M^2 (1 \pm A) x^*_1\right]\,,
\end{align}%
where $\rho^*_{0}=\rho_0/\rho_I$, $p^*_0=p_0/(\rho_I g\lambda)$, with $(1-A)$ for $x^*_1<0$ (i.e., the lighter fluid) and $(1+A)$ for $x^*_1>0$ (i.e., the heavier fluid). The resulting initial background stratifications are shown for a variety of Mach numbers in Figure \ref{f:init}, where the size of the density difference at $x^*_1=0$ is determined by the value of $A$ ($A=0.04$ in the present study).  

%______________________________
%______________________________

Following the procedure extensively outlined by Reckinger \emph{et al.} \cite{Reckinger2016}, a single-mode velocity perturbation was applied at $t=0$ to initialize the RTI. Although they are not perfect representations of multi-mode engineering problems found in ICF and other practical applications, the present single-mode simulations can nevertheless be used to gain insights into compressibility-driven physics and dynamics. As shown by Reckinger \emph{et al.} \cite{Reckinger2016}, single-mode simulations can expose any numerical directional bias in the code, which is generally hidden in multi-mode simulations. As a result, single-mode simulations allow the opportunity to ensure that the simulations are completely resolved from the initial state through to late times, and also allow simple checks for symmetry and the introduction of extraneous perturbation modes throughout the simulation. In addition, the results of Wei \& Livescu \cite{Wei2012} show that single-mode RTI may represent the upper bound for the multi-mode growth rate at low Atwood numbers, when the Reynolds number is sufficiently large. 

%%%%%%%%%%%%%%%%%%%%%%%%%
%% Numerical Method
%%%%%%%%%%%%%%%%%%%%%%%%%
\section{Details of the Direct Numerical Simulations\label{sec:method}}
Due to the spatial localization of the developing region, the RTI problem lends itself naturally to state-of-the-art adaptive grid numerical methods. In particular, to effectively capture the instability evolution, very long domains are needed to ensure that late-time growth is captured, but very small grid spacing is required to fully resolve the high gradients at the interface of the instability. For a static computational grid with fixed cell size, this results in a very dense grid and incredibly high computational costs. During the majority of the simulation, however, very fine grid resolutions far away from the interface are unnecessary and, as a result, high grid compression ratios can be achieved through the use of adaptive grid approaches. A method that has proven effective at achieving high compression ratios is the Parallel Adaptive Wavelet Collocation Method (PAWCM) \cite{Reckinger2010,Reckinger2016}, which is the method that is applied here.

%=====================================
% Wavelet-Based Grid Adaptation
%=====================================
\subsection{Wavelet-Based Grid Adaptation}
The PAWCM numerical approach has been applied previously to the simulation of compressible RTI by Reckinger \emph{et al.} \cite{Reckinger2016}, where validation and details of the numerical method are exhaustively outlined. These details are repeated only briefly here, and the reader is referred to \cite{Reckinger2016} for additional information. 

Fundamentally, PAWCM uses the natural properties of the wavelet transform to locate areas of steep gradients and to provide direct control over the grid cell size used to resolve the gradients. Essentially, through PAWCM, a flow field variable is transformed into wavelet space, resulting in wavelet basis functions and coefficients that are localized in both wave and physical spaces. From there, the coefficients are passed through a thresholding filter where all of the coefficients with magnitudes above the parameter $\varepsilon$ are kept, and any of those below $\varepsilon$ are set to zero. The resulting thresholded decomposition can thus be written for a generic variable $f$ as
\begin{equation}
  \label{e:wavelet}
 f_{\geq}(x) = \sum_{k} c^0_k \phi^0_k(x) + \mathop{{\sum_{j=0}^{\infty} \sum_{\alpha=1}^{2^n-1} {\sum_l}}}_{\left|d^{\alpha,j}_l\right| \geq \varepsilon \left|\left| f \right|\right| } d^{\alpha,j}_l \psi^{\alpha,j}_l(x)\,,
\end{equation}%
where $\phi_k$ are scaling functions on the coarsest level, $c_k$ are the corresponding coarse-level wavelet coefficients, $\psi_l$ are the scaling interpolating functions on any arbitrary level, $d_l$ are the coefficients to which the thresholding is applied, $l$ and $k$ represent physical grid points, and $\alpha$ and $j$ represent the wavelet family and level of resolution, respectively \cite{Vasilyev2000a,Nejadmalayeri2015}. The effect of setting any one of the coefficients $d_l$ to zero is the removal of a grid point at that level of resolution. These coefficients take on large values for large gradients, and small values in relatively uniform regions. The effective resolution is set by a base grid size and the limit put on $j$ (referred to as $j_{max}$ herein). This results in the error being $\mathcal{O}(\varepsilon)$ and the resolution in a single direction being $p\cdot2^{(j_{max} - 1)} $, where $p$ is the base resolution \cite{Vasilyev2000a, Schneider2010a,Nejadmalayeri2015}.

As outlined in Reckinger \emph{et al.} \cite{Reckinger2016}, PAWCM has been implemented in a way that enables it to work with finite difference approaches to solving governing equations such as those in Eqs.\ (\ref{e:continuity})-(\ref{e:species}). In solving these equations, fourth-order central differences have been applied spatially, and a third-order total variation diminishing explicit Runge-Kutta scheme has been applied in time. The PAWCM algorithm is highly parallelized, having successfully run on up to 5,000 cores, and is able to perform arbitrary domain decompositions using the Zoltan library. It has a tree-like data structure for easy MPI communications, as well as direct error control. As a result, the additional computational overhead introduced by the wavelet methodology is offset by the capability to use many processors and to achieve grid compression ratios greater than 90\% \cite{Vasilyev2000a, Schneider2010a,Nejadmalayeri2015}.

Substantial discussion was provided in Reckinger \emph{et al.} \cite{Reckinger2016} regarding the flow variables on which to adapt the grid in the DNS. Since the wavelet method is so flexible, it is possible to adapt the grid on any flow field variable that is calculable and of interest. In the present study, adaptation for the initial time steps was performed using the vorticity, the norm of the strain rate tensor, and the gradient of the species mass fraction $Y_2$, in addition to the velocity and mass fraction fields. This approach allowed the RTI to develop with sufficient accuracy prior to further refining the grid on more complex flow variables at later times to reflect the increasing complexity of the flow. In particular, at late times in the present study, adaptation was performed using the baroclinic torque to ensure that this dynamically important term was fully resolved for the analysis of the vorticity dynamics. Additional details on grid convergence and resolution can be found in \cite{Reckinger2016}. 

%=====================================
% Simulation Setup
%=====================================
\subsection{Simulation Setup}
In the present study, PAWCM is used to solve the governing equations outlined in Section \ref{subsec:goveqs} for the background and initial conditions described in Section \ref{subsec:rti}. The simulations have been carried out in 2D and the total domain size was $16\lambda$ in the $x_1$ direction and $\lambda$ in the $x_2$ direction, where $\lambda$ corresponds to the wavelength of the applied perturbation. The maximum effective grid resolution resulting from the adaptive wavelet approach was $\Delta x^*= 2.4\times 10^{-4}$, where $\Delta x^*=\Delta x/\lambda$ and $\Delta x$ is the grid cell size. 

The Atwood number studied was 0.04, and the Mach numbers used were 0.3 (nearly incompressible), 0.6, 0.9, and 1.2. The perturbation Reynolds numbers, $Re_p$, investigated were 5,000, 10,000, and 20,000, giving a total of twelve simulations performed in the present study (i.e., four different values of $M$, and three values of $Re_p$ for each $M$). The highest Reynolds number is of particular interest because it has been shown to be the minimum perturbation Reynolds number necessary to reach the chaotic growth regime for the incompressible limit of this particular case \cite{Wei2012}. Each of the simulations were performed up to a non-dimensional time of $t^*=t/\sqrt{\lambda/ g}=20$, corresponding to the time at which the bubble and spike had reached heights of roughly $\lambda$ (or $x_1^*=\pm 1$) for the $M=0.3$ case. 

Boundaries in the $x_2$ direction are taken to be periodic. In the $x_1$ direction, at the top and bottom of the domain, shear-free slip boundary conditions were implemented with numerical diffusion buffer zones immediately before each boundary interior to the domain. The purpose of these ``open'' boundary conditions is to essentially mimic an infinite domain and to ensure that both the background stratification is preserved and that none of the shocks introduced by the RTI initialization are reflected back into the domain. In particular, the buffer zones ensure that any shockwaves are dissipated prior to reaching the boundaries \cite{Reckinger2016}.

As discussed in \cite{Reckinger2016}, some artificial thickening of the interface at $x^*_1=0$ and $t^*=0$ can be beneficial since the thicker interface can act as a buffer layer to absorb other numerical errors. In general, however, thicker interfaces have the potential to introduce asymmetries in the initial conditions which propagate as undesirable longer-time asymmetries during RTI growth. Based on these two competing considerations, the number of points across the interface was chosen to be 16, to both minimize the asymmetry and to gain some measure of beneficial buffering effects. Finally, it was found that higher resolutions led to better initial conditions. At a level of $j_{max} = 7$, it was found that the asymmetry drops below machine precision, and thus this level of resolution was deemed sufficient for the present simulations.

%%%%%%%%%%%%%%%%%%%%%%%%%
%% Evolution of Rayleigh Taylor Instabilities
%%%%%%%%%%%%%%%%%%%%%%%%%
\section{Rayleigh Taylor Instability Growth and Characteristics\label{sec:results}}
The PAWCM-enabled simulations performed here are designed to allow examination of stratification strength (as parameterized by $M$) and Reynolds number (as parameterized by $Re_p$) effects on RTI growth and characteristics. In the following, these two effects are investigated with a primary focus on the heights and velocities of bubbles and spikes formed during the RTI development. Here the ``height'' is denoted $h$ and refers to the absolute value of the respective distances from $x_1=0$ of the bubble and spike ``tips'' in the $x_1$ direction. The bubble and spike tips correspond to the $99\%$ and $1\%$ mass fraction values, respectively. Bubble and spike velocities, denoted $u_h$, are computed from the time derivatives of the bubble and spike heights. An analysis of the dynamics underlying the observed bubble and spike behaviors is outlined in Section \ref{sec:vorticity}. 

%______________________________
\begin{figure}[t]
\centering
 \includegraphics[scale=0.9]{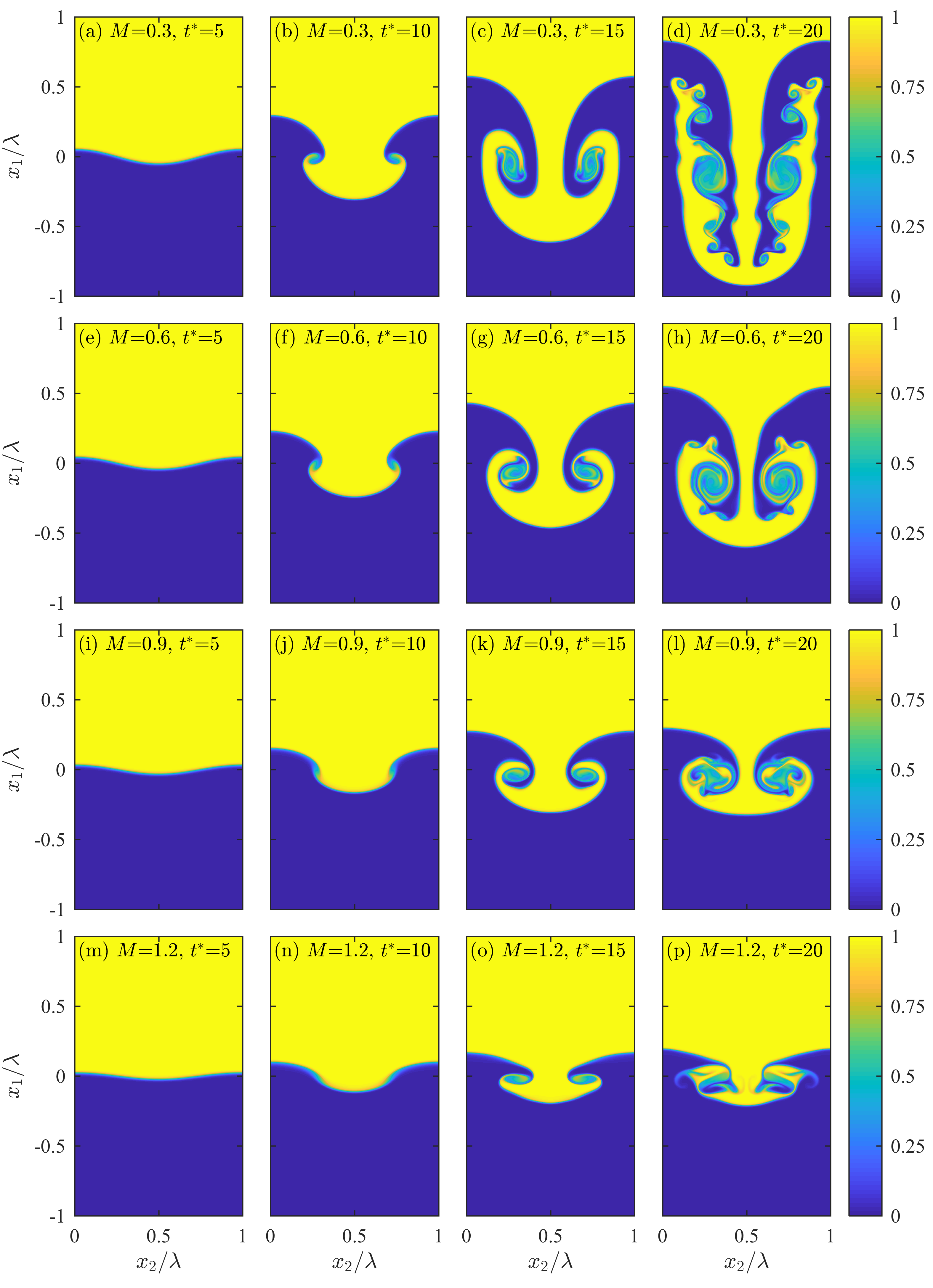}
\caption{[Color online] Instantaneous fields of the heavier species mass fraction, $Y_2$, in $x_1$-$x_2$ planes as a function of non-dimensional time $t^*=t\sqrt{g/\lambda}$ for stratification strengths $M=0.3$, 0.6, 0.9, and 1.2 (increasing from top to bottom). The progression in time from $t^*=5$ to $t^*=20$ is shown in columns from left to right.}
\label{f:ma_comp}
\end{figure}
%______________________________

%______________________________
\begin{figure}[t]
\centering
 \includegraphics[scale=0.9]{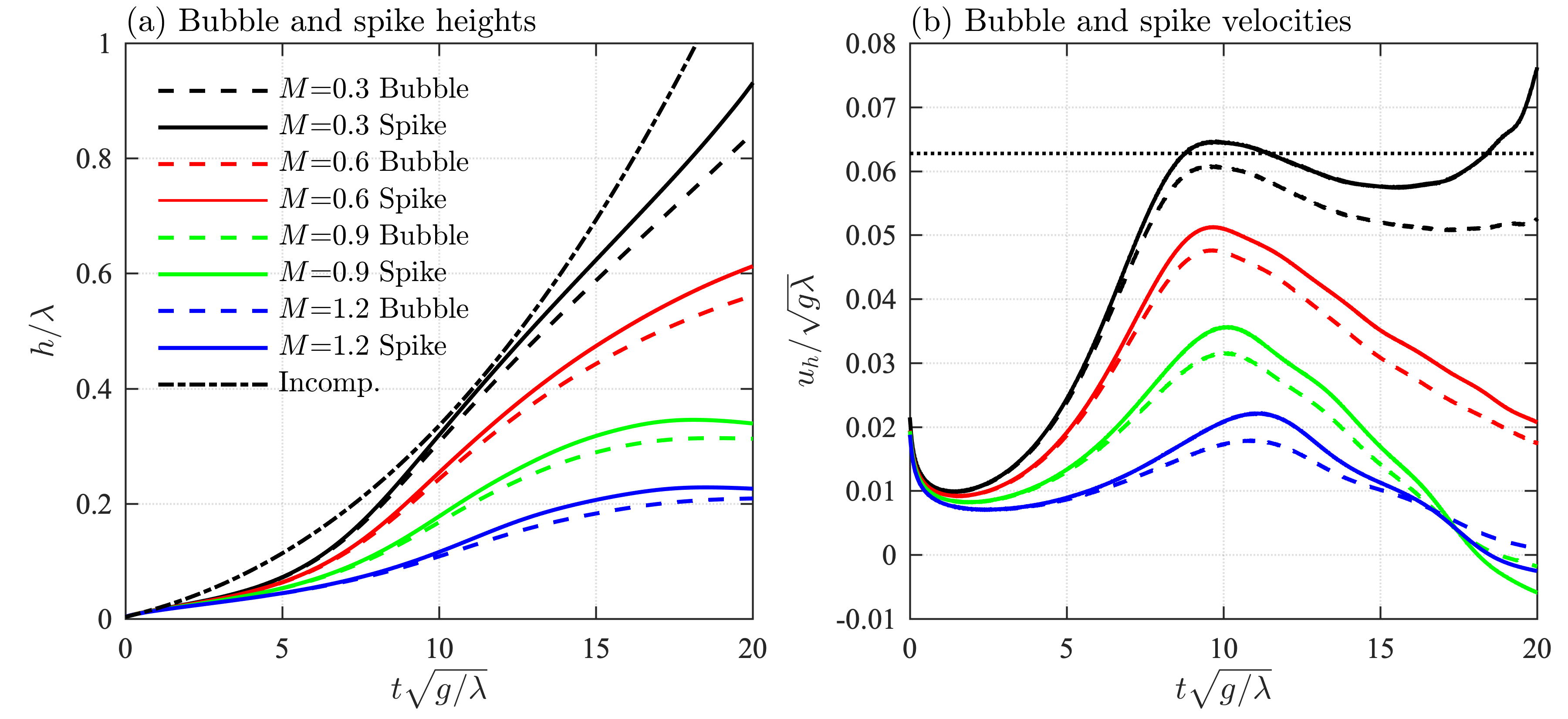}
 \caption{ \label{f:hvbs_ma}[Color online] Time series of bubble and spike tip heights, $h$, (panel a) and velocities, $u_h$, (panel b) for $M=0.3$, 0.6, 0.9, and 1.2. Bubble results are shown by dashed lines and spike results are shown by solid lines. Heights, velocities, and times have each been non-dimensionalized using $\lambda$ and $g$. The dash-dot line in panel (a) shows incompressible results from Wei \& Livescu \cite{Wei2012} and the horizontal dotted line in panel (b) shows the predicted bubble velocity from drag and potential flow models \cite{Birkhoff1960,Oron2001,Goncharov2002}, $u_h/\sqrt{g\lambda}\approx 0.063$.}
\end{figure}
%______________________________

%=====================================
% Effects of Stratification Strength
%=====================================
\subsection{Effects of Stratification Strength}
Figure \ref{f:ma_comp} shows RTI growth as a function of time for each of the four stratification strengths, where $Re_p=20,000$ in all cases. For each case, bubbles and spikes form soon after initializing the simulation and the RTI grows as $t$ increases. Small-scale features in each case become increasingly pronounced as the RTI evolves, and secondary vortices are most prominent for the weakest stratification (i.e., $M=0.3$). The corresponding bubble and spike growths decrease as the stratification strength increases; for the strongest stratification (i.e., $M=1.2$), the RTI growth is halted relatively early in its evolution. 

Consistent with the fields in Figure \ref{f:ma_comp}, Figure \ref{f:hvbs_ma}(a) shows that the suppression of RTI growth compared to the incompressible case from Wei \& Livescu \cite{Wei2012} occurs for all stratifications considered. For the two strongest stratifications (i.e., $M=0.9$ and 1.2), the bubble and spike each reach maximum heights before $t^*=20$ and stop growing. 

The dependence of RTI growth on stratification strength can be investigated further by considering time series of the bubble and spike tip velocities, as shown in  Figure \ref{f:hvbs_ma}(b). This figure indicates that bubble and spike velocities for the larger Mach numbers all trend towards zero, indicative of the complete suppression of RTI for strong stratifications. For $M=0.3$, however, there is a re-acceleration of the spike tip shortly after $t^*=15$. 

In addition to these changes in the bubble and spike heights with varying stratification strength, Figure \ref{f:hvbs_ma}(a) also shows that spikes reach consistently greater heights than bubbles for all $M$. This asymmetry, particularly for low $M$, is not present in the purely incompressible case of Wei \& Livescu \cite{Wei2012}, where it was found that for the low Atwood number case of 0.04, bubble and spike heights were close until after the re-acceleration regime. As indicated by Figure \ref{f:hvbs_ma}(b), the velocities at the tips of the spikes are consistently larger than those at the tips of the bubbles, although the difference between these velocities becomes significant only for the $M=0.3$ case after 
$t^*>10$. 

It should be noted that full suppression of RTI growth for all but the lowest value of $M$ cannot be predicted based solely on considerations of the potential energy of the system. This is shown in Figure \ref{f:hvbs_ma}(b), where only the lowest value of $M$ reaches a plateau near the velocity predicted from either drag \cite{Birkhoff1960,Oron2001} or potential flow \cite{Goncharov2002} models (namely, $u_h/\sqrt{g\lambda}\approx 0.063$). 

Based on the DNS results for $M=0.3$ to 1.2, the primary observations are that larger stratifications are associated with decreasing bubble and spike growth rates, resulting in a suppression of the RTI for all but the smallest value of $M$ studied here, and that smaller stratifications are associated with more asymmetric bubble and spike growth rates. This amounts to an anomalous asymmetry at low stratifications (i.e. $M=0.3$), since both zero and large $M$ limits are more symmetrical. The results concerning the suppression of the instability are in general agreement with those from prior studies \cite{Lafay2007,Gauthier2010,Wei2012,Reckinger2016,Gauthier2017}. To better understand the dynamics leading to RTI suppression and the development of bubble and spike asymmetries, an analysis of the underlying vorticity dynamics is performed in Section \ref{sec:vorticity}.

%______________________________
\begin{figure}[t]
\centering
 \includegraphics[scale=0.9]{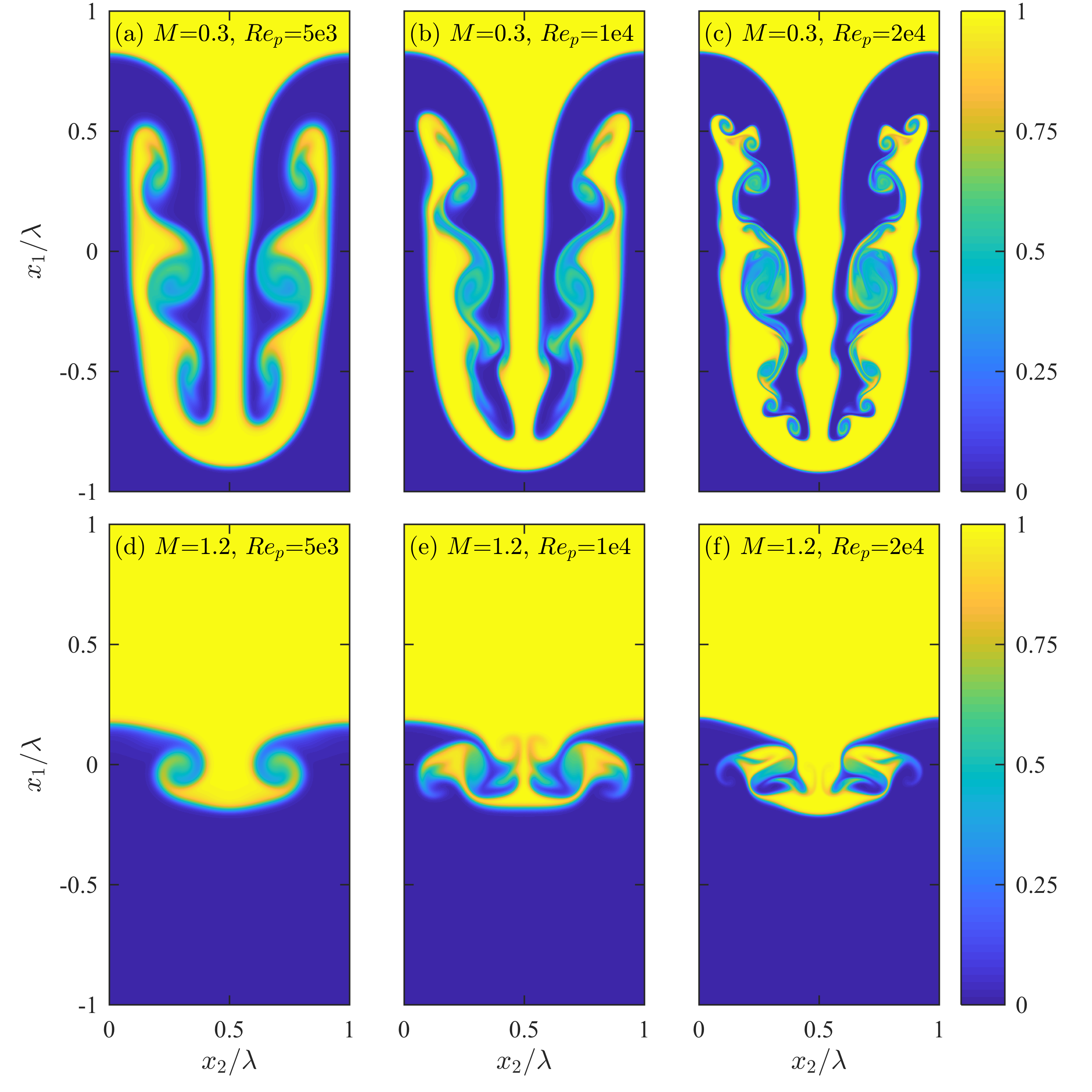}
 \caption{[Color online] Instantaneous fields of the heavier species mass fraction, $Y_2$, in $x_1$-$x_2$ planes for Reynolds numbers $Re_p=5\times 10^3$, $1\times 10^4$, and $2\times 10^4$ (left to right columns; denoted `5e3', `1e4', and `2e4', respectively), for stratification strengths $M=0.3$ (top row) and $M=1.2$ (bottom row). Results are shown at $t^*=t\sqrt{g/\lambda}=20$ in each case.}
\label{f:rey_comp}
\end{figure}
%______________________________

%______________________________
\begin{figure}[t]
\centering
 \includegraphics[scale=0.9]{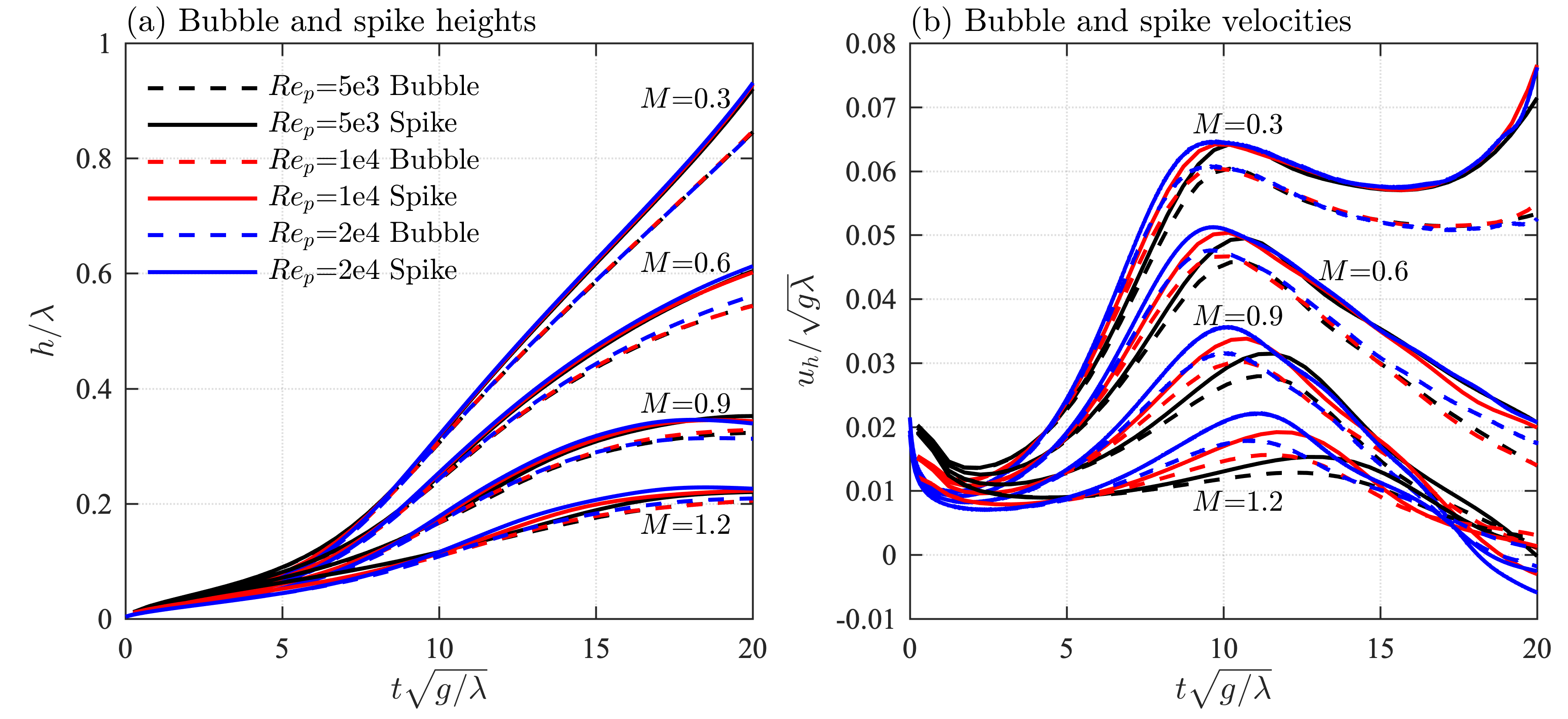}
 \caption{ \label{f:hvbs_ma_re}[Color online] Time series of bubble and spike tip heights, $h$, (panel a) and velocities, $u_h$, (panel b) for $M=0.3$, 0.6, 0.9, and 1.2, at $Re_p=5\times 10^3$, $1\times 10^4$, and $2\times 10^4$ (denoted `5e3', `1e4', and `2e4', respectively). Bubble results are shown by dashed lines and spike results are shown by solid lines. Results for $Re_p=5\times 10^3$, $1\times 10^4$, and $2\times 10^4$ are shown using black, red, and blue lines, respectively. Heights, velocities, and times have each been non-dimensionalized using $\lambda$ and $g$.}
\end{figure}
%______________________________

%=====================================
% Effects of Reynolds Number
%=====================================
\subsection{Effects of Reynolds Number}
Figure \ref{f:rey_comp} shows RTI growth for the weakest (i.e., $M=0.3$) and strongest (i.e., $M=1.2$) stratifications for perturbation Reynolds numbers $Re_p=5,000$, $10,000$, and $20,000$. There is little qualitative dependence of the bubble and spike heights on $Re_p$, indicating that these large-scale characteristics of RTI growth are already in an asymptotic limit for $Re_p=5,000$. This is consistent with the results from Wei \& Livescu \cite{Wei2012}, where it was found that there is little difference in the RTI growth rates before the onset of the very late chaotic development for values of $Re_p$ above roughly $1,500$. 

Despite the relative similarity of the large-scale structure for the three values of $Re_p$ examined here, however, there is substantial dependence of small-scale structure on $Re_p$. In particular, Figure \ref{f:rey_comp} shows that an increasing amount of small-scale detail emerges as $Re_p$ increases, corresponding to the occurrence of viscous dissipation at increasingly smaller scales. This increase in scale range with increasing $Re_p$ results in the formation of secondary vortices for $M=0.3$. Even though there is also increasing small scale structure for $M=1.2$ with increasing $Re_p$, the formation of secondary vortices is less pronounced for this higher stratification due to the overall suppression in the RTI growth.

From a quantitative perspective, Figure \ref{f:hvbs_ma_re} shows bubble and spike heights and velocities for each of the four values of $M$ examined in the present study. For the bubble and spike heights shown in Figure \ref{f:hvbs_ma_re}(a), there is little or no dependence on $Re_p$ for any stratification strength. However, for the velocities in Figure \ref{f:hvbs_ma_re}(b), there is a clear trend towards faster initial accelerations as $Re_p$ increases. The bubbles and spikes also reach larger maximum velocities as $Re_p$ increases. However, at very early times in the evolution for each $M$,  during diffusive growth, bubble and spike velocities are largest for small $Re_p$, eventually crossing over in each case at $t^*\approx 5$ such that the higher $Re_p$ cases have greater velocities at later times. This result is consistent with the crossover in speeds observed by Wei \& Livescu \cite{Wei2012} and, to a somewhat lesser extent, by Gauthier \cite{Gauthier2017}.

These trends are consistent for all stratification strengths, although the differences with $Re_p$ become more pronounced as $M$ increases. For example, the peak bubble and spike velocities for $M=1.2$ are reached at roughly $t^*=12$ when $Re_p=20,000$ and at roughly $t^*=14$ when $Re_p=5,000$. After reaching the peak values, however, the bubble and spike velocities become substantially less dependent on $Re_p$. For the case with smallest $M$, the results approach the nearly incompressible limit where, as shown by  Wei \& Livescu \cite{Wei2012}, no dependence on $Re_p$ is observed above $Re_p\approx 1,500$ during the times examined here (before the onset of late time chaotic development regime). 

Taken together, these results indicate that, for the values of $Re_p$ examined here, there is little dependence of the global RTI growth on $Re_p$ during the later stages of the instability at higher stratifications and through the early re-acceleration stage for $M=0.3$. However, the early time evolution, small scale structure, and the appearance of secondary vortices are all substantially affected by $Re_p$. Given the increasing effect of $Re_p$ with increasing $M$, it may be the case that $Re_p$ effects become increasingly pronounced for even stronger stratifications than the $M=1.2$ case examined here; exploring such more strongly stratified scenarios is left as a direction for future research. 

%%%%%%%%%%%%%%%%%%%%%%%%%
%% Vorticity Dynamics
%%%%%%%%%%%%%%%%%%%%%%%%%
\section{Vorticity Dynamics for Compressible Rayleigh Taylor Instability\label{sec:vorticity}}
Properties and dynamics of the vorticity vector, $\omega_i = \epsilon_{ijk}\partial u_k/\partial x_j$, where $\epsilon_{ijk}$ is the alternating tensor, have been widely studied to understand flow behavior in a variety of contexts. For compressible flows more specifically, vorticity has been studied in shock-driven \cite{Larsson2009,Larsson2013,Livescu2016}, reacting \cite{Steinberg2009,Hamlington2011}, and various types of buoyant \cite{Gauthier2008} flows, revealing the dynamical importance of variable density effects such as dilatation and baroclinic torque. In the case of RTI, however, only Gauthier \cite{Gauthier2017} has examined vorticity dynamics in the fully compressible regime, and for only one value of the initial stratification strength.

%______________________________
\begin{figure}[t]
\centering
\includegraphics[scale=0.9]{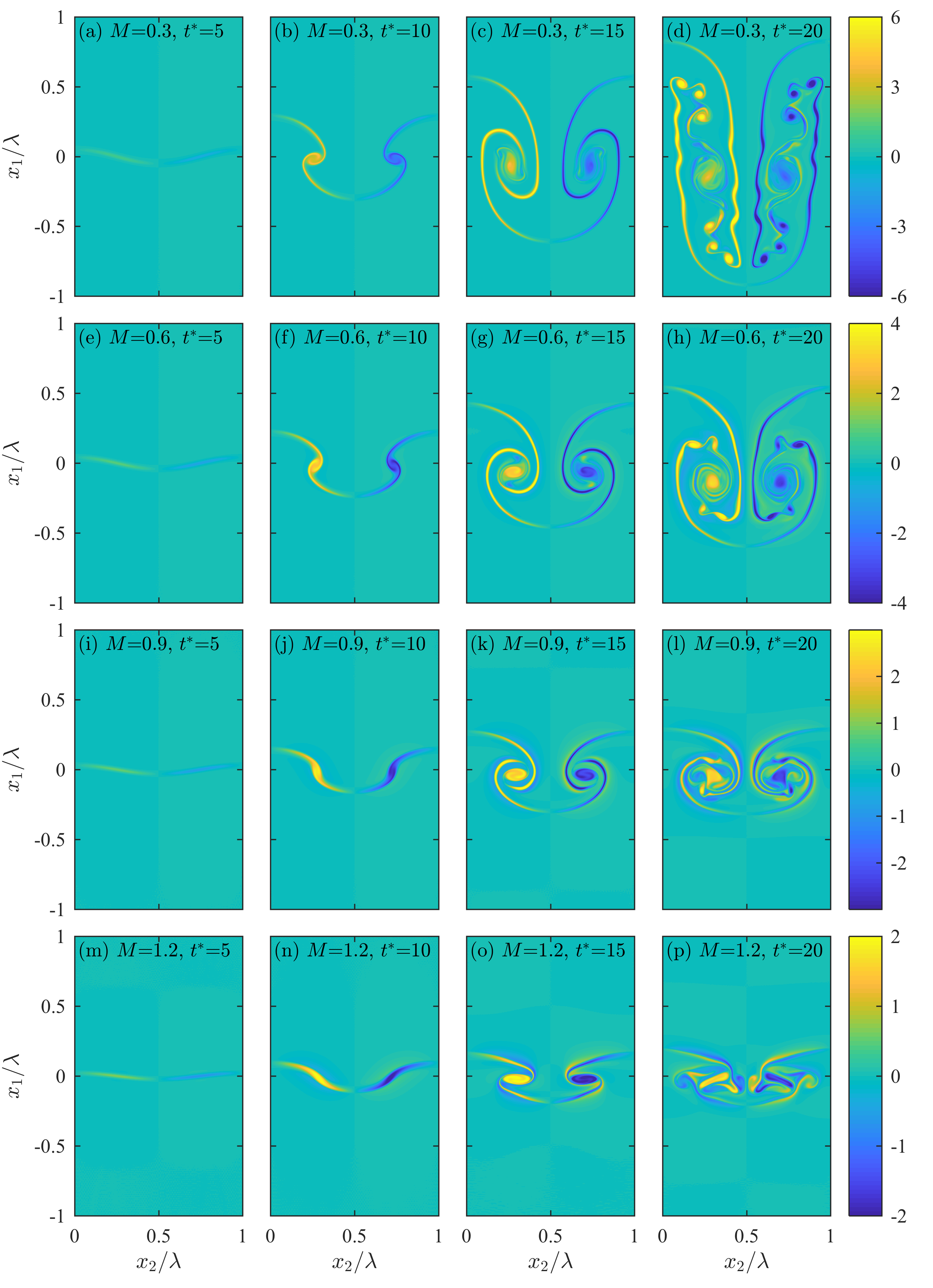}
\caption{[Color online] Instantaneous fields of the non-dimensional vorticity $\omega_3^*=\omega_3\sqrt{\lambda/g}$, in $x_1$-$x_2$ planes as a function of non-dimensional time $t^*=t\sqrt{g/\lambda}$ for stratification strengths $M=0.3$, 0.6, 0.9, and 1.2 (increasing from top to bottom). The progression in time from $t^*=5$ to $t^*=20$ is shown in columns from left to right.}
\label{f:vort_m3m12}
\end{figure}
%______________________________

In the following sections, properties of the vorticity during RTI growth are outlined as a function of stratification strength, and terms in the non-dimensional compressible vorticity transport equation are subsequently examined to understand the underlying dynamics. The role of baroclinic torque, in particular, in the suppression of RTI growth for strong stratifications and in the formation of bubble and spike asymmetries for weak stratifications is outlined. It should be noted that the importance of baroclinic torque in RTI growth is not new or surprising and has been highlighted in several previous studies \cite{Lafay2007,Schneider2016a,Gauthier2013,Gauthier2017,Wei2012}. The primary contribution of the current work is in explaining how the baroclinic torque varies with initial stratification strength, as well as how RTI suppression and asymmetry arise from a dynamical perspective. 
%=====================================
% Description of Vorticity Evolution
%=====================================
\subsection{Vorticity Evolution for Compressible RTI}
In the 2D simulations, $\omega_3^*$ is the only nonzero component of the vorticity, and Figure \ref{f:vort_m3m12} shows the temporal evolution of this component for each of the stratification strengths. For each stratification strength, the vorticity field initially develops as a vortex pair with generally positive vorticity for $x_2<0.5$ and negative vorticity for $x_2>0.5$. These initial vortex pairs evolve by moving downwards slowly in the domain, while the Kelvin-Helmholtz instability on the sides of the bubbles and spikes sheds further vortex pairs. The overall spatial extent of vorticity production is greatest for weak stratification (i.e., $M=0.3$), with ``fronts'' of non-zero vorticity magnitude that propagate upwards and downwards in an analogous way to the propagation of bubbles and spikes, respectively, as shown in Figure \ref{f:ma_comp}. The vorticity evolution at $M=0.3$ is reminiscent of the overall picture in the incompressible case, with induced vortical velocity helping the instability growth and leading to re-acceleration and late time chaotic development. However, at higher $M$ values, no additional vortex pairs are generated. The overall magnitude of the vorticity is also shown in Figure \ref{f:vort_m3m12} to decrease with increasing $M$.

\begin{figure}[t]
\centering
 \includegraphics[scale=0.9]{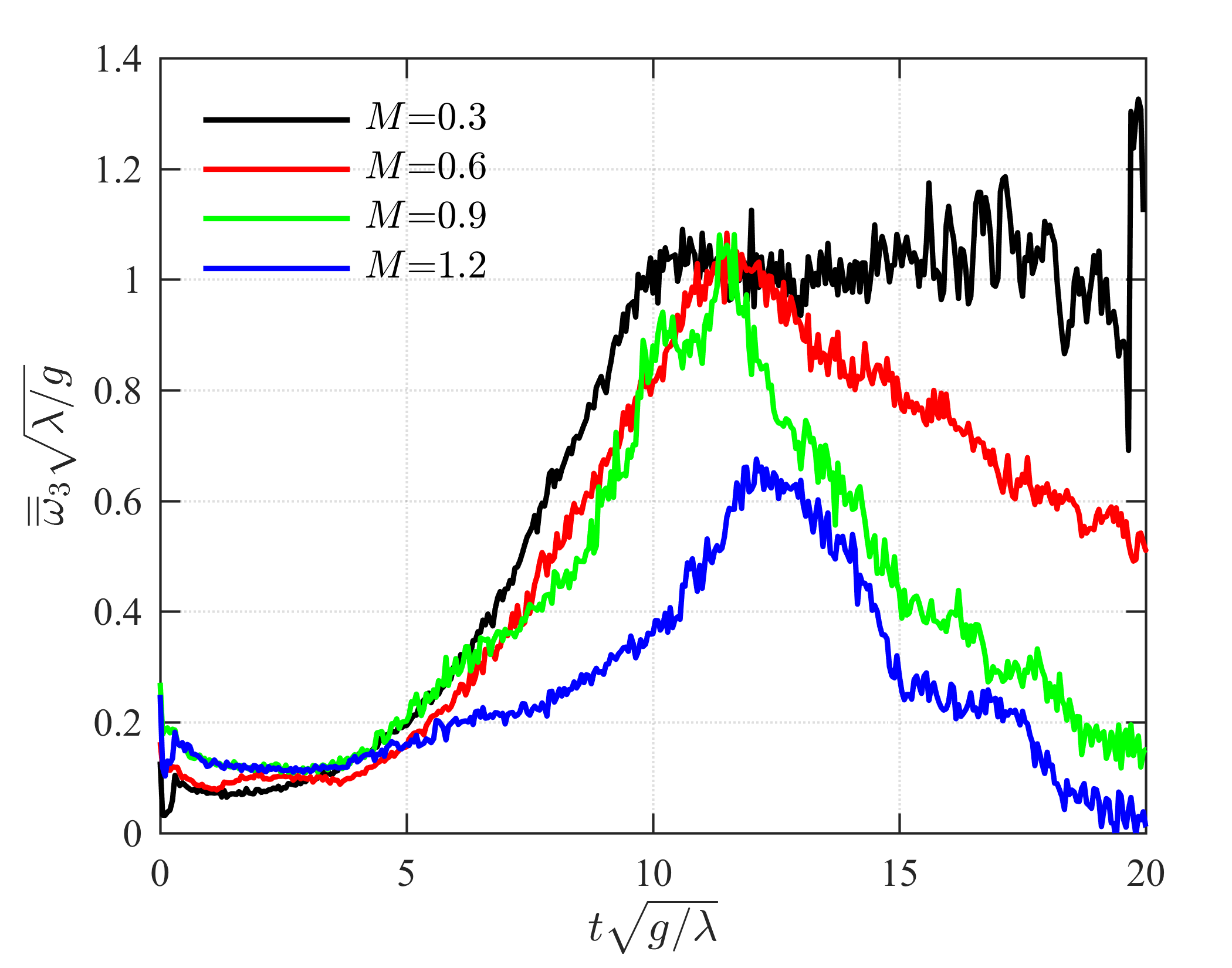}
 \caption{ \label{f:avgvort}[Color online] Temporal evolution of the average vorticity $\overline{\overline{\omega}}_3$ over the left half of the domain (i.e., $x_2<\lambda/2$) for stratification strengths $M=0.3$, 0.6, 0.9, and 1.2, where the averaging operator is defined in Eq.\ (\ref{e:fullavg}).}
\end{figure}

The overall $M$-dependence of the vorticity magnitude is also explored in Figure \ref{f:avgvort} using the vorticity averaged over the half domain, denoted $\overline{\overline{\omega}}_3$, where the half-domain averaging operator is defined for an arbitrary quantity $f$ as
\begin{equation}\label{e:fullavg}
\overline{\overline{f}}(t) = \frac{1}{\lambda^2}\int_0^{\lambda/2} \int_{-\lambda}^\lambda  f(x_1,x_2,t) dx_1 dx_2\,.
\end{equation}
Figure \ref{f:avgvort} shows that $\overline{\overline{\omega}}_3$ generally increases at early times at a rate that is larger with decreasing stratification. After the initial growth of $\overline{\overline{\omega}}_3$ shown in Figure \ref{f:avgvort}, the average vorticity decreases with time for all but the weakest stratification (i.e., $M=0.3$). This result mirrors the suppression of RTI growth for all but the weakest stratification, seen in Figure \ref{f:hvbs_ma}. 

%______________________________
%______________________________

%=====================================
% Non-Dimensional Vorticity Tranport Equation
%=====================================
\subsection{Non-Dimensional Compressible Vorticity Transport Equation}
The dynamics governing vorticity evolution in compressible RTI can be understood from the non-dimensional vorticity transport equation, which reveals the explicit dependence of the dynamics on $A$, $M$, and $Re_p$. A similar equation was derived using the Boussinesq approximation by Schneider \& Gauthier \cite{Schneider2016a}, although any explicit dependence on the initial stratification strength was omitted in the derivation. Here, the non-dimensional transport equation is derived for the fully compressible case, permitting the explicit identification of dependencies on stratification strength $M$. 

By taking the curl of the momentum equation in Eq.\ (\ref{e:navier}), the transport equation for the 3D vorticity vector is obtained for a variable density, variable viscosity compressible flow as
\begin{equation}\label{e:vorticity}
\frac{D \omega_i}{D t} = \omega_j S_{ij} - \omega_i S_{kk} - \epsilon_{ijk}\frac{\partial v}{\partial x_j}\frac{\partial p}{\partial x_k} + \epsilon_{ijk}\frac{\partial}{\partial x_j}\left[v\frac{\partial (2\mu S'_{kl})}{\partial x_l}\right]\,,
\end{equation}
where $D/Dt\equiv \partial/\partial t + u_i \partial/\partial x_i$ is the Lagrangian derivative and $v\equiv 1/\rho$ is the specific volume, which is used here instead of $\rho$ to simplify the derivation. The first term on the right-hand side of Eq.\ (\ref{e:vorticity}) represents vortex stretching, the second term represents dilatation, the third term is the baroclinic torque, and the last term is viscous diffusion, where the viscous stress tensor $\tau_{kl}$ has been expressed in terms of the deviatoric strain rate tensor $S'_{kl}$ [see Eq.\ (\ref{e:tau})]. 

The last term in Eq.\ (\ref{e:vorticity}), representing viscous diffusion, can be separated into an essentially incompressible term that is present for all $M$, regardless of whether viscosity $\mu$ is spatially and temporally varying, and into a term that is only present when $\mu$ is non-constant. In the present simulations, $\mu=\nu \rho$, where $\nu$ is a constant given in terms of problem parameters as in Eq.\ (\ref{e:rey}), and $\rho$ is the spatially and temporally varying density. Expansion of the diffusion term in Eq.\ (\ref{e:vorticity}) then gives the vorticity transport equation for a variable density, variable viscosity flow as
\begin{equation}\label{e:vorticity2}
\frac{D \omega_i}{D t} = \omega_j S_{ij} + \nu \frac{\partial^2 \omega_i}{\partial x_j \partial x_j}- \omega_i S_{kk} - \epsilon_{ijk}\frac{\partial v}{\partial x_j}\frac{\partial p}{\partial x_k} - 2\nu \epsilon_{ijk}\frac{\partial}{\partial x_j}\left(\frac{S'_{kl}}{v}\frac{\partial v}{\partial x_l}\right)\,.
\end{equation}
The first two terms on the right-hand side of this equation are present even in constant density, constant viscosity flows, while the last three terms are only nonzero when $v$ (and, by extension, the density) is non-constant. 

Using the characteristic time scale $\sqrt{\lambda/g}$ to define the non-dimensional vorticity $\omega_i^*\equiv \omega_i\sqrt{\lambda/g}$, and using $\rho_I g\lambda$ as the characteristic pressure, Eq.\ (\ref{e:vorticity2}) can be written in non-dimensional form as
\begin{equation}\label{e:vortnondim}
\frac{D \omega^*_i}{D t^*} = \omega^*_j S^*_{ij} +\sqrt{\frac{A}{1+A}}\frac{1}{Re_p}\frac{\partial^2 \omega^*_i}{\partial x^*_j \partial x^*_j} - \omega^*_i S^*_{kk}- \epsilon_{ijk}\frac{\partial v^*}{\partial x^*_j}\frac{\partial p^*}{\partial x^*_k}-\sqrt{\frac{A}{1+A}}\frac{2}{Re_p}\epsilon_{ijk}\frac{\partial}{\partial x^*_j}\left[S'^*_{kl}\frac{\partial (\ln v^*)}{\partial x^*_l}\right]\,.
\end{equation}
Based on the above equation, both diffusive terms thus scale in an identical fashion with $A$ and $Re_p$. It should be noted, however, that the stratification strength $M$ does not appear explicitly in Eq.\ (\ref{e:vortnondim}), although it is present implicitly in the baroclinic torque term [i.e., the third term on the right-hand side of Eq.\ (\ref{e:vortnondim})]. To reveal this dependence, this term can be rewritten by defining new perturbation variables $v'^*$ and $p'^*$ that express $v^*$ and $p^*$ relative to their respective $A=0$ background stratifications as
\begin{equation}\label{e:decomp}
v'^*(\textbf{x}^*,t^*) \equiv v^*(\textbf{x}^*,t^*)-\left[v^*_{M}(x_1^*)-1\right]\,,\quad p'^*(\textbf{x}^*,t^*) \equiv p^*(\textbf{x}^*,t^*)-\left[p^*_{M}(x_1^*)+x_1^*- \frac{1}{M^2}\right] \,,
\end{equation}
where $\textbf{x}^*=[x^*_1,x^*_2,x^*_3]$, $v_M^*\equiv 1/\rho_M^*$, and $\rho_M^*$ and $p_M^*$ correspond to the $A=0$ profiles of $\rho^*_0$ and $p_0^*$ from Eqs.\ (\ref{e:rhoothermfinal}) and (\ref{e:pisothermfinal}), respectively. The $A=0$ profiles are used for normalization purposes to avoid discontinuities in the derivatives of the background profiles that arise when $A$ is nonzero (particularly for the first derivative of $p_0^*$). The resulting $A=0$ profiles are, nevertheless, not substantially different than the $A=0.04$ profiles (see Figure \ref{f:init}) and serve the purpose of explicitly revealing the dependence of the baroclinic torque on $M$. 

\begin{figure}[t]
\centering
 \includegraphics[scale=0.9]{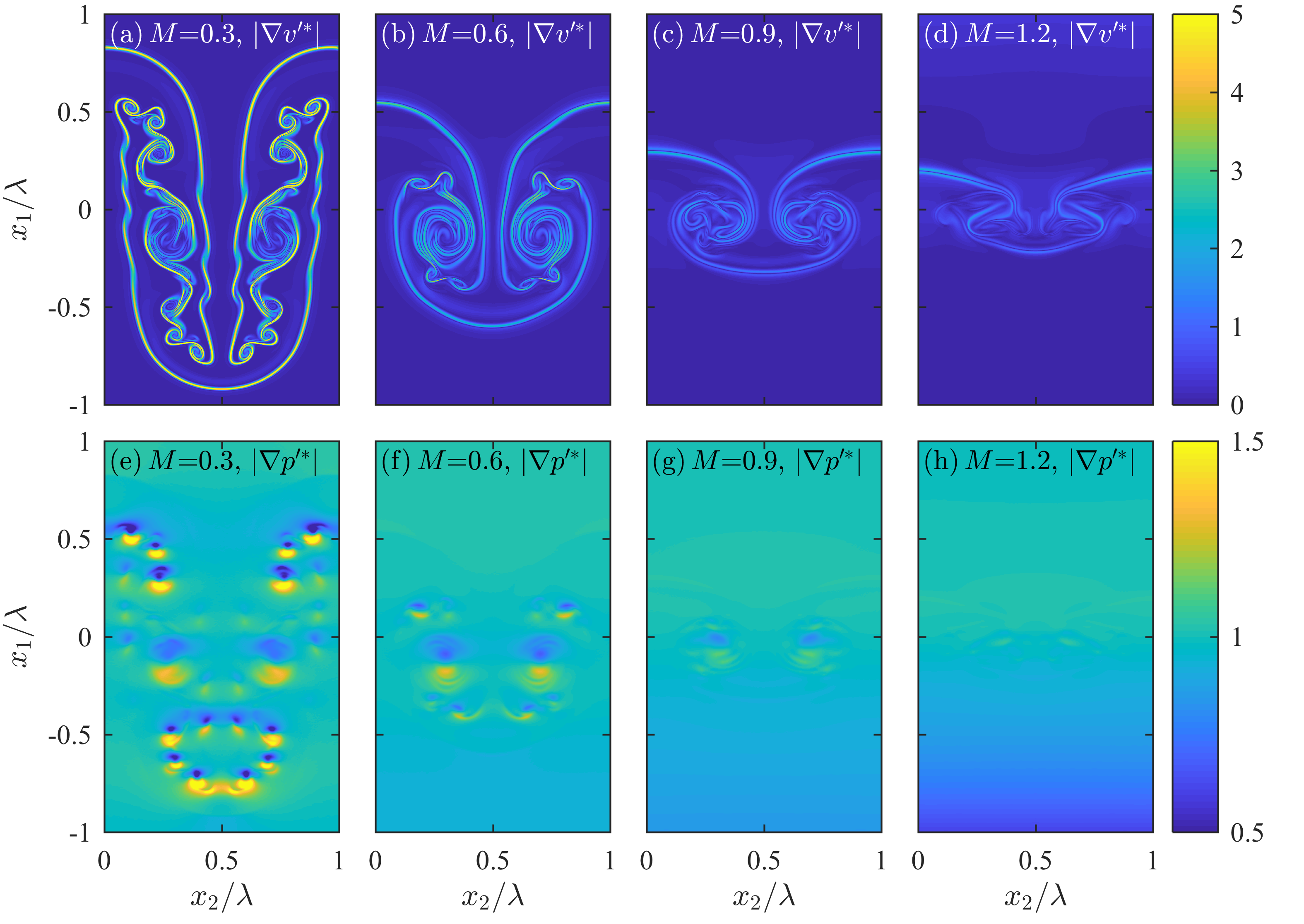}
\caption{[Color online] Instantaneous fields showing the magnitudes of $\partial v'^*/\partial x^*_i$ (top row) and $\partial p'^*/\partial x^*_i$ (bottom row), where the perturbation gradients are given in Eqs.\ (\ref{e:decomp1}) and (\ref{e:decomp2}). Fields are shown at non-dimensional time $t^*=t\sqrt{g/\lambda}=20$ and for stratification strengths $M=0.3$, 0.6, 0.9, and 1.2 (left to right).}
\label{f:grads}
\end{figure}

The decompositions in Eq.\ (\ref{e:decomp}) are designed to yield $v'^*=v^*$ and $p'^*=p^*$ in the limit as $M\rightarrow 0$, as well as $\partial v'^*/\partial x^*_i= \partial v^*/\partial x^*_i$ and $\partial p'^*/\partial x^*_i= \partial p^*/\partial x^*_i$ in the same limit. The baroclinic torque term in Eq.\ (\ref{e:vortnondim}) depends only on gradients of $v^*$ and $p^*$, and the equivalency of the perturbation and total gradients can be shown for $M\rightarrow 0$ as  
\begin{align}
\frac{\partial v'^*}{\partial x_i^*}= \frac{\partial v^*}{\partial x_i} -M^2 v_M^*\delta_{i1} \quad &\Rightarrow \quad \frac{\partial v'^*}{\partial x_i^*}= \frac{\partial v^*}{\partial x_i} \textrm{ as } M\rightarrow 0\,, \label{e:decomp1}\\
\frac{\partial p'^*}{\partial x_i^*}= \frac{\partial p^*}{\partial x_i} +(M^2 p_M^*-1)\delta_{i1} \quad &\Rightarrow \quad \frac{\partial p'^*}{\partial x_i^*}= \frac{\partial p^*}{\partial x_i} \textrm{ as } M\rightarrow 0\,,\label{e:decomp2}\
\end{align} 
where $v_M^*\rightarrow 1$ and $M^2 p_M^*\rightarrow 1$ as $M\rightarrow 0$. As $M$ becomes large and background stratification becomes increasingly strong, the magnitude of $\partial v'^*/\partial x^*_i$ becomes increasingly small and $\partial p'^*/\partial x^*_i$ approaches the background stratification everywhere, as shown in Figure \ref{f:grads}.

%______________________________
%______________________________

Using Eqs.\ (\ref{e:decomp1}) and (\ref{e:decomp2}), it can be shown that the baroclinic torque on the right-hand side of Eq.\ (\ref{e:vortnondim}) can be written as
\begin{equation}\label{e:expand}
 - \epsilon_{ijk}\frac{\partial v^*}{\partial x^*_j}\frac{\partial p^*}{\partial x^*_k}=  - \epsilon_{ijk}\frac{\partial v'^*}{\partial x^*_j}\frac{\partial p'^*}{\partial x^*_k} + \left(M^2 p^*_M-1\right) \epsilon_{ij1}\frac{\partial v'^*}{\partial x^*_j}+ \left( M^2 v^*_M\right) \epsilon_{ij1}\frac{\partial p'^*}{\partial x^*_j}\,,
\end{equation}
where it is assumed that $v_M^*$ and $p^*_M$ depend only on $x_1^*$. The first term on the right in Eq.\ (\ref{e:expand}) represents the baroclinic torque that is independent of the initial background stratification, and this is the only remaining term in the limit as $M\rightarrow 0$. The second and third terms represent the baroclinic torques associated with the initial stratified background pressure and specific volume fields, respectively. It should be noted that the present analysis is specific to the isothermal forms for $v_M^*$ and $p^*_M$ obtained from Eqs.\ (\ref{e:decomp1}) and (\ref{e:decomp2}), and that the scaling may differ for different initial background conditions (e.g., isentropic or isobaric conditions).  

After substituting Eq.\ (\ref{e:expand}) into Eq.\ (\ref{e:vortnondim}), the non-dimensional 3D vorticity transport equation is obtained for a compressible flow with initial background stratification as
\begin{align}\label{e:vortnondimfinal}
\frac{D \omega^*_i}{D t^*} = &\omega^*_j S^*_{ij}+\sqrt{\frac{A}{1+A}}\frac{1}{Re_p}\frac{\partial^2 \omega^*_i}{\partial x^*_j \partial x^*_j}- \omega^*_i S^*_{kk} -  \epsilon_{ijk}\frac{\partial v'^*}{\partial x^*_j}\frac{\partial p'^*}{\partial x^*_k} \\ &+ \left(M^2 p^*_M-1\right) \epsilon_{ij1}\frac{\partial v'^*}{\partial x^*_j}+ \left( M^2 v^*_M\right) \epsilon_{ij1}\frac{\partial p'^*}{\partial x^*_j}-\sqrt{\frac{A}{1+A}}\frac{2}{Re_p}\epsilon_{ijk}\frac{\partial}{\partial x^*_j}\left[S'^*_{kl}\frac{\partial (\ln v^*)}{\partial x^*_l}\right]\,,\nonumber
\end{align}
where, once more, the first four terms are present even in the limit as $M\rightarrow 0$ and the fifth and sixth terms are only significant for nonzero $M$. The corresponding transport equation for the vorticity magnitude $\omega^*\equiv (\omega_i^* \omega_i^*)^{1/2}$ is given by
\begin{align}\label{e:vortnondimfinalmag0}
\frac{D \omega^*}{D t^*} = &\overbrace{\widehat{\omega}^*_i\omega^*_j S^*_{ij}}^{\mathcal{T}^*_1} \overbrace{+\sqrt{\frac{A}{1+A}}\frac{\widehat{\omega}^*_i}{Re_p}\frac{\partial^2 \omega^*_i}{\partial x^*_j \partial x^*_j}}^{\mathcal{T}^*_2}\overbrace{- \omega^* S^*_{kk}}^{\mathcal{T}^*_3} \overbrace{- \widehat{\omega}^*_i\epsilon_{ijk}\frac{\partial v'^*}{\partial x^*_j}\frac{\partial p'^*}{\partial x^*_k}}^{\mathcal{T}^*_4} \\ &\underbrace{+ \left(M^2 p^*_M-1\right) \widehat{\omega}^*_i\epsilon_{ij1}\frac{\partial v'^*}{\partial x^*_j}}_{\mathcal{T}^*_5}\underbrace{+ \left( M^2 v^*_M\right) \widehat{\omega}^*_i\epsilon_{ij1}\frac{\partial p'^*}{\partial x^*_j}}_{\mathcal{T}^*_6}\underbrace{-\sqrt{\frac{A}{1+A}}\frac{2\widehat{\omega}^*_i}{Re_p}\epsilon_{ijk}\frac{\partial}{\partial x^*_j}\left[S'^*_{kl}\frac{\partial (\ln v^*)}{\partial x^*_l}\right]}_{\mathcal{T}^*_7}\,.\nonumber
\end{align}
where $\widehat{\omega}^*_i \equiv \omega^*_i/\omega^*$ is the vorticity unit vector (where the magnitude of $\widehat{\omega}^*_i$ is unity by definition). This expression is valid for any $M$, $A$, and $Re_p$ provided that $v_M^*$ and $p_M^*$ are given by Eqs.\ (\ref{e:decomp1}) and (\ref{e:decomp2}) and that $\nu$ is constant. In the above expression, $\mathcal{T}^*_1$ represents production and destruction of $\omega^*$ due to vortex stretching, $\mathcal{T}^*_2$ represents diffusion of vorticity by constant viscosity, $\mathcal{T}^*_3$ represents dilatational effects, $\mathcal{T}^*_4$ represents stratification-independent baroclinic torque, $\mathcal{T}^*_5$ represents baroclinic torque production associated with the background pressure field, $\mathcal{T}^*_6$ represents baroclinic torque production associated with the background specific volume field, and $\mathcal{T}^*_7$ represents diffusion associated with variable viscosity. In the limit as $M\rightarrow 0$, both $\mathcal{T}^*_6$ and $\mathcal{T}^*_7$ terms go to zero.

%______________________________
\begin{figure}[t]
\centering
 \includegraphics[scale=0.9]{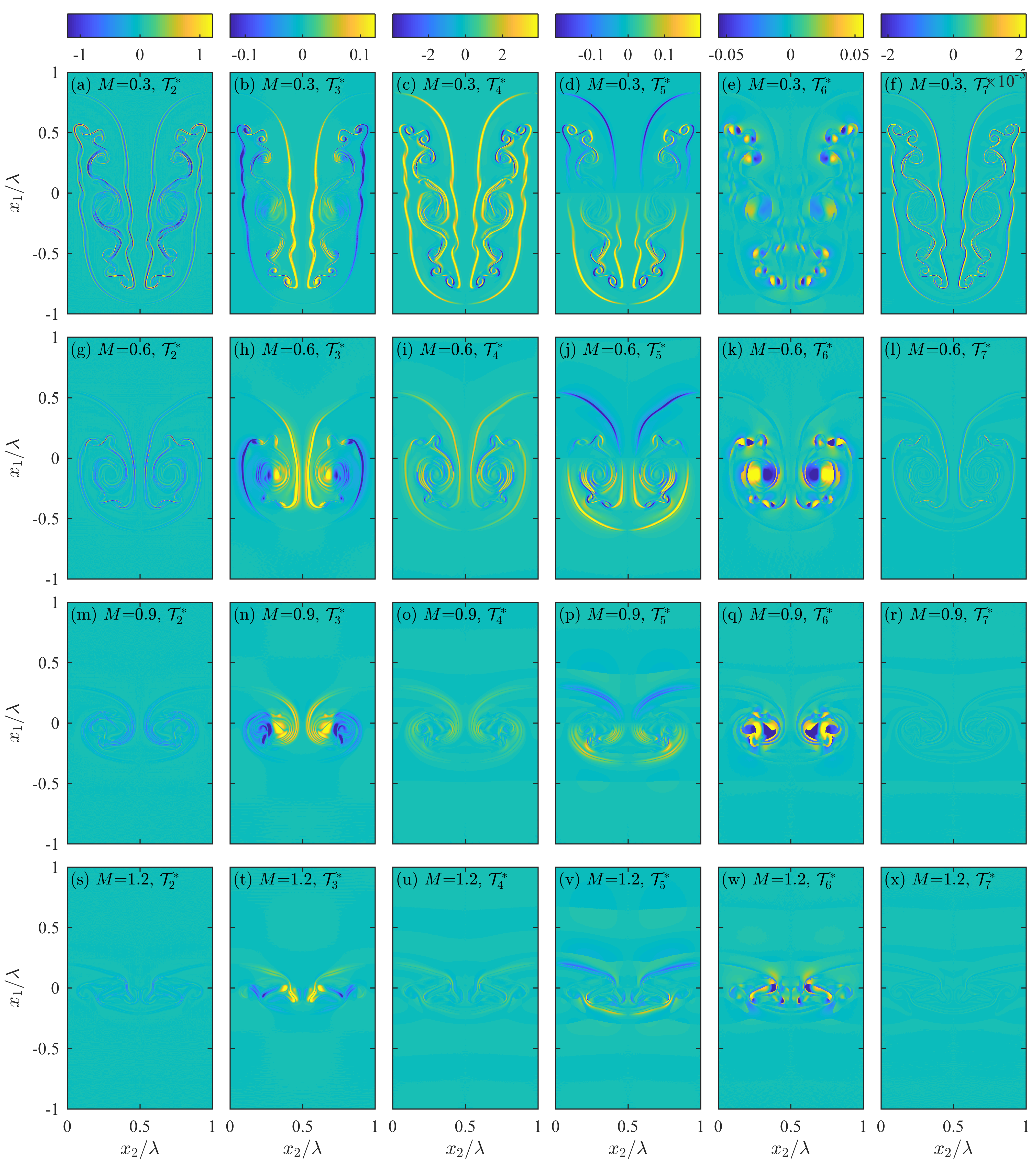}
\caption{[Color online] Instantaneous fields of $\mathcal{T}^*_i=\mathcal{T}_i(\lambda/g)$ appearing in Eq.\ (\ref{e:vortnondimfinalmag0}), which describes the dynamics of $\omega^*=|\omega^*_3|$, for the 2D simulation cases. Fields are shown at non-dimensional time $t^*=t\sqrt{g/\lambda}=20$ for terms $\mathcal{T}^*_2$--$\mathcal{T}^*_7$ (left to right) and for stratification strengths $M=0.3$, 0.6, 0.9, and 1.2 (top to bottom).}
\label{f:t1t2}
\end{figure}
%______________________________

%=====================================
% Effects of stratification
%=====================================
\subsection{Effects of Stratification Strength on the Dynamics of the Vorticity}
Figure \ref{f:t1t2} shows fields of $\mathcal{T}^*_2$-$\mathcal{T}^*_7$ from Eq.\ (\ref{e:vortnondimfinalmag0}) for the four different stratification strengths (with $Re_p = 20,000$ in all cases) at a late stage ($t^*=20$) in the 2D simulations. The vortex stretching term $\mathcal{T}^*_1$ is identically zero in 2D and is thus not shown here. 

For small $M$, Figure \ref{f:t1t2} shows that $\mathcal{T}^*_4$, representing the perturbation baroclinic torque, is primarily positive (indicating vorticity production) and roughly an order of magnitude larger than $\mathcal{T}^*_5$ (baroclinic torque due to the background pressure) and $\mathcal{T}^*_6$ (baroclinic torque due to the background specific volume). The dilatation term $\mathcal{T}^*_3$ has a similar magnitude to $\mathcal{T}^*_5$ and $\mathcal{T}^*_6$, but is smaller than $\mathcal{T}^*_4$. The constant viscosity diffusion term,  $\mathcal{T}^*_2$, is much larger than the variable viscosity contribution, $\mathcal{T}^*_7$, and reaches peak magnitudes similar to, but still smaller than, $\mathcal{T}^*_4$. 

\begin{figure}[t]
\centering
 \includegraphics[scale=0.9]{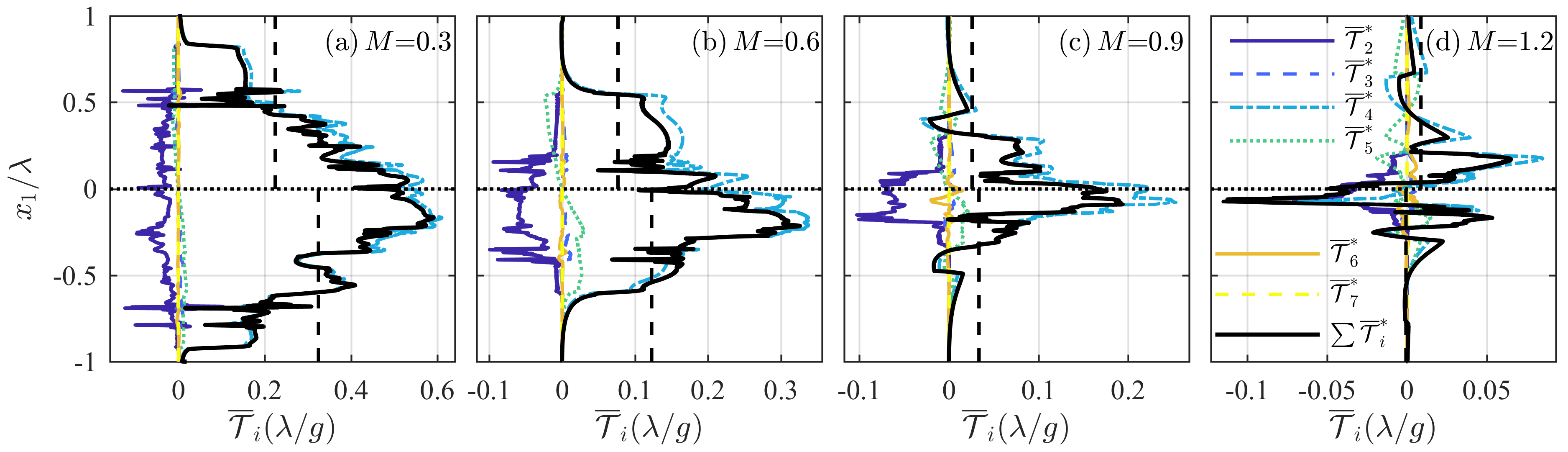}
\caption{[Color online] Spatial dependence along the $x_1$ direction of the half-domain averages of $\mathcal{T}^*_2$--$\mathcal{T}^*_7$ appearing in Eq.\ (\ref{e:vortnondimfinalmag0}) for stratification strengths $M=0.3$, 0.6, 0.9, and 1.2 (a-d) in the 2D simulation cases. The sums of terms $\mathcal{T}^*_2$--$\mathcal{T}^*_7$ are also shown. The averaging operator $\overline{(\cdot)}$ is defined in Eq.\ (\ref{e:halfavg}) and $\overline{\mathcal{T}}_i$ is written in non-dimensional form as $\overline{\mathcal{T}}_i^*=\overline{\mathcal{T}}_i(\lambda/g)$. The vertical dashed lines show averages of the sum of all terms for $x_1>0$ and $x_1<0$. All results are shown at a non-dimensional time of $t^*=t\sqrt{g/\lambda}=20$.}
\label{f:Tmach}
\end{figure}

Figure \ref{f:t1t2} thus indicates that the primary dynamical effects for low $M$ are the perturbation baroclinic torque (i.e., $\mathcal{T}^*_4$) and constant viscosity diffusion (i.e., $\mathcal{T}^*_2$), although the former dominates the latter, resulting in the growth of the instability for low $M$. The relative magnitudes of these terms are shown in Figure \ref{f:Tmach}, where the terms $\mathcal{T}^*_i$ from Eq.\ (\ref{e:vortnondimfinalmag0}) are averaged over half of the domain along the $x_2$ direction to give $\overline{\mathcal{T}}^*_i$ as a function of $x_1$, with the average defined as
\begin{equation}\label{e:halfavg}
\overline{f}(x_1,t) = \frac{2}{\lambda}\int_0^{\lambda/2} f(x_1,x_2,t) dx_2\,.
\end{equation}
Figure \ref{f:Tmach} also shows results for the averages of $\overline{\mathcal{T}}^*_i$ over $x_1$ for $x_1<0$ and $x_1>0$. For the weakest stratification examined here, Figure \ref{f:Tmach}(a) shows that the enstrophy is created on average due almost entirely to the perturbation baroclinic torque. There is only a relatively small enstrophy destruction contribution due to the constant viscosity diffusion.

Although the perturbation baroclinic torque $\mathcal{T}^*_4$ can become locally negative due to density inversions (i.e. negative density gradients) created by vortical motions, Figures \ref{f:t1t2} and \ref{f:Tmach} show that this term remains mostly positive for all but the strongest stratification, due to the presence of the instability. Nevertheless, $\mathcal{T}^*_4$ does decrease in magnitude as $M$ increases and, in particular, Figure \ref{f:Tmach}(d) shows that this term can contribute to the destruction of vorticity magnitude for sufficiently large stratification. This is consistent with Figure \ref{f:grads}, which shows that $\partial v'^*/\partial x^*_i$ approaches zero, while $\partial p'^*/\partial x^*_i$ becomes close to 1, as $M$ increases. The reduced vorticity production at larger stratifications corresponds to the suppression of the instability growth.

For all values of $M$ considered here, Figures \ref{f:t1t2} and \ref{f:Tmach} show that $\mathcal{T}^*_4$ has the largest magnitude in general, but, since it decreases with $M$, becomes more similar in magnitude to the other terms at the largest stratification considered. At $M=1.2$, the vorticity production is much smaller, consistent with the overall suppression of the instability. For large stratifications, the reduced vorticity magnitude also translates into lower self-propagating velocity for the vortex pairs generated at the bubble/spike interface. In turn, this results in part of the fresh fluid brought towards the bubble/spike peaks by the induced vortical velocity returning back to the mixing layer. Thus, at $M=0.9$, Figure \ref{f:Tmach}(c) shows density inversions (i.e., negative $\overline{\mathcal{T}}^*_4$ or stabilizing regions) near the edges of the layer, while at $M=1.2$ in Figure \ref{f:Tmach}(d) these regions can occur throughout the layer.

%______________________________
%______________________________

Variations in the magnitudes of $\mathcal{T}^*_5$ and $\mathcal{T}^*_6$ with $M$ shown in Figures \ref{f:t1t2} and \ref{f:Tmach} are somewhat more complicated. In particular, the peak magnitudes of both terms increase from $M=0.3$, but start decreasing again at larger $M$ and become smaller for $M=1.2$. Term $\mathcal{T}^*_5$ reaches its peak magnitude at slightly smaller Mach number than $\mathcal{T}^*_6$ ($M\sim0.6$ versus $M\sim 0.9$). At large $M$, the gradient contributions to both terms become uniform, with values of 0 and 1, respectively (see Figure \ref{f:grads}). The prefactors $(M^2 p_M^*-1)$ and $M^2 v_M^*$ depend on the initial background stratification and become large in the far-field, but are small near the centerline, even for large $M$. Therefore, as the instability growth is suppressed at large stratifications, $\mathcal{T}^*_5$ and $\mathcal{T}^*_6$ are more confined to the region close to the centerline and never reach regions with large prefactor values. 

%______________________________
\begin{figure}[t]
\centering
 \includegraphics[scale=0.9]{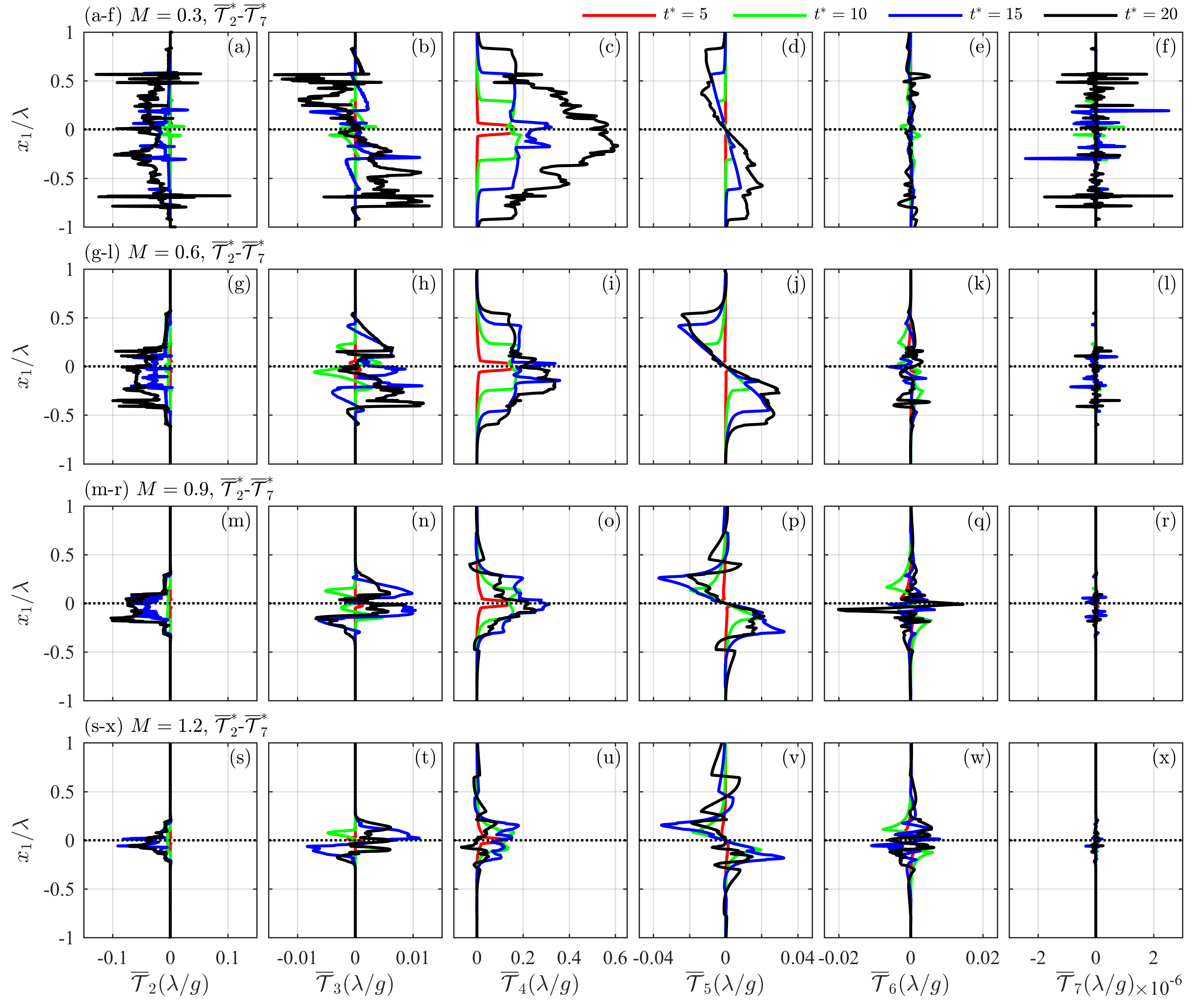}
\caption{[Color online] Spatial dependence of the half-domain averages of $\mathcal{T}^*_2$--$\mathcal{T}^*_7$ (columns from left to right) appearing in Eq.\ (\ref{e:vortnondimfinalmag0}) for stratification strengths $M=0.3$ (a-f), $M=0.6$ (g-l), $M=0.9$ (m-r), and $M=1.2$ (s-x) at non-dimensional times $t^*=t\sqrt{g/\lambda}=5$, 10, 15, and 20 (red, green, blue, and black lines, respectively). The averaging operator $\overline{(\cdot)}$ is defined in Eq.\ (\ref{e:halfavg}) and $\overline{\mathcal{T}}_i$ is written in non-dimensional form as $\overline{\mathcal{T}}_i^*=\overline{\mathcal{T}}_i(\lambda/g)$.}
\label{f:Tevol}
\end{figure}
%______________________________

Perhaps most significantly, both $\mathcal{T}^*_5$ and $\mathcal{T}^*_6$ exhibit asymmetries that affect the overall growth of the instability. Aside from local inversions, Figures \ref{f:t1t2} and \ref{f:Tmach} show that $\mathcal{T}^*_5$ presents a top-bottom asymmetry with respect to the $x_1=0$ initial location of the instability (i.e., it is positive on the spike side and negative on the bubble side). On the other hand, Figure \ref{f:t1t2} shows that $\mathcal{T}^*_6$ presents a left-right asymmetry with respect to the interface between the heavy and light fluid, with negative values inside the spike and positive values inside the bubble regions. Conversely, the dilatation term $\mathcal{T}^*_3$ shows the opposite left-right asymmetry, with positive values inside the spike and negative values inside the bubble regions. It should be noted, however, that $\mathcal{T}^*_5$ becomes \emph{larger} with respect to $\mathcal{T}^*_4$ as $M$ increases. The term $\mathcal{T}^*_4$ is itself also asymmetric, as shown in Figure \ref{f:Tmach}, and it is likely that this asymmetry is the underlying cause of the differences in bubble and spike growth rates, particularly for small $M$. Moreover, weak asymmetry in bubble and spike growth rates is observed in the incompressible limit, indicating that $\mathcal{T}^*_5$ may be a contributor to, but not the sole cause of, the asymmetry, since this term approaches zero as $M\rightarrow 0$. 

As explained above, the bubble-spike asymmetry is small in the incompressible case before the chaotic stage, it becomes noticeable at $M=0.3$, and then decreases again at large stratifications. The history of the top-down asymmetry in the vorticity generation can also be seen from the time evolutions of $\overline{\mathcal{T}}_i$ in Figure \ref{f:Tevol}. This figure does not identify the left-right asymmetry, which has a more dynamical effect, as it influences the vortical motions separately within the bubble and spike regions. However, it does show that $\overline{\mathcal{T}}^*_4$ begins symmetrical and develops the top-down asymmetry at some later time. On the other hand, $\overline{\mathcal{T}}^*_5$  is asymmetric from the beginning, such that it represents the source of this asymmetry. This is consistent with the incompressible flow results, where terms $\overline{\mathcal{T}}^*_3$, $\overline{\mathcal{T}}^*_5$, and $\overline{\mathcal{T}}^*_6$ are zero, and $\overline{\mathcal{T}}^*_4$ remains relatively symmetrical until later times. Again, at large stratifications, the overall reduction in vorticity production and suppression of the instability prevents the bubble/spike asymmetry from becoming more pronounced. 

Finally, Figures \ref{f:t1t2}--\ref{f:Tevol} show that the variable viscosity diffusion term, $\mathcal{T}^*_7$, is negligible for all $M$ considered here and all times. By contrast, the magnitude of the constant viscosity diffusion term, $\mathcal{T}^*_2$, remains more uniform with increasing $M$, although it does become more consistently negative as $M$ increases, as shown most clearly in Figure \ref{f:Tevol}. This indicates that constant viscosity diffusion is the primary term leading to destruction of vorticity magnitude, and this term begins to rival the magnitude of the perturbation baroclinic torque term (i.e., $\mathcal{T}^*_4$) for large $M$. 

Taken together, these results are indicative of larger vorticity production within the spike region, as compared to the bubble region, due to compressibility and stratification effects. Because the bubble and spike vertical axes are maintained throughout the flow evolution for the single mode case, the vorticity field itself retains a similar symmetry. This results in an induced vortical velocity along the bubble/spike axes, which helps the instability grow, similar to the incompressible case \cite{Wei2012}. However, for the compressible case, the dilatation term and baroclinic contributions sum up to a bubble/spike asymmetry even at low Mach numbers. At higher Mach numbers, due to the overall suppression of the instability, these contributions also decrease and the asymmetry becomes small again. Overall, the primary dynamical balance is between constant viscosity diffusion, which leads to the destruction of vorticity magnitude, and perturbation baroclinic torque, which leads to vorticity magnitude production.  

%%%%%%%%%%%%%%%%%%%%%%%%%
%% Conclusions
%%%%%%%%%%%%%%%%%%%%%%%%%
\section{Conclusions\label{sec:conclusions}}
In the present study, wavelet-based adaptive mesh refinement has been used to perform DNS of 2D single-mode compressible low Atwood number RTI for four different isothermal stratification strengths, corresponding to Mach numbers from 0.3 to 1.2, and for three different perturbation Reynolds numbers from $5,000$ to $20,000$. The simulation results have been examined to understand the effects of stratification strength and Reynolds number on the characteristics, dynamics, and rate of RTI growth. In the present context, compressibility is controlled through the values of the background pressure at the interface between the heavier and lighter fluids, which also affects the background stratification strength, and would be considered flow, as opposed to fluid, compressibility. In this context, the incompressible limit is reached as the speed of sounds goes to infinity by increasing the interface pressure and temperature, such that the interface density remains constant. The practical setup corresponds to an enclosed fluid system that is uniformly heated (i.e., heating at constant volume). 

For weak stratifications, RTI growth was found to undergo a re-acceleration after reaching a plateau in the growth rate that approximately matched predictions from potential flow theory. As the stratification strength increased, however, this re-acceleration was found to no longer occur, and the RTI growth was suppressed; this suppression occurred in the present study for all Mach numbers greater than 0.3. For weak stratifications, the bubble was found to grow at a slower rate than the spike, but this asymmetry progressively weakened as the stratification strength increased. The Reynolds number was found to have little impact on RTI growth for the range of Mach numbers and for the simulation length examined here. However, small-scale structure was found to become more pronounced as the Reynolds number increased. At very early times, during the diffusive stage, the growth rates were larger at smaller Reynolds numbers, but the instability became faster during the linear and weakly nonlinear stages at higher Reynolds numbers, consistent with prior studies of Reynolds number effects \cite{Wei2012,Gauthier2017}.

To determine the origins of the observed results, the dynamics of the vorticity magnitude were examined in detail. A non-dimensional compressible vorticity transport equation was derived to explicitly show dependencies on the Mach, Atwood, and Reynolds numbers, and the effects of stratification strength were studied for each of the terms in the transport equation. This analysis showed that incompressible baroclinic torque was the dominant driver of RTI growth for the range of stratifications considered, and its decrease at higher stratifications corresponded to the overall instability suppression. Asymmetries in the RTI growth were found to be the result of compressibility effects, as a consequence of the dilatation term and background stratification contributions to the baroclinic torque. However, for strong stratifications, since the instability did not evolve far from the centerline, the latter contributions remain small and the bubble/spike asymmetry does not become pronounced. 

In total, the simulations and analysis performed in this study have enabled the three questions posed in Section \ref{sec:intro} to be fully addressed. However, much work remains to be done. In particular, the present analysis of vorticity dynamics should be extended to multi-mode initial perturbations, to different stratification types (e.g., isopycnic and isentropic stratifications), and to 3D cases where vortex stretching effects in the vorticity dynamics are nonzero. It would also be of interest to explore longer simulation times for the weakly stratified cases to determine whether the chaotic development regime noted by Wei \& Livescu \cite{Wei2012} is recovered in the context of fully compressible simulations. 

%%%%%%%%%%%%%%%%%%%%%%%%%
%% Acknowledgements
%%%%%%%%%%%%%%%%%%%%%%%%%
\section{Acknowledgements}
This work was made possible in part by funding from the LDRD program at Los Alamos National Laboratory through project number 20150568ER. SAW was supported by Los Alamos National Laboratory, under Grant No. 316898. Computational resources were provided by the LANL Institutional Computing (IC) Program. Dr.\ Oleg Vasilyev provided inspiration and early direction for this work. 

%%%%%%%%%%%%%%%%%%%%%%%%%
%% References
%%%%%%%%%%%%%%%%%%%%%%%%%
%\bibliographystyle{unsrt}
%\bibliography{crti}

\end{document}